\documentclass[ reprint,
superscriptaddress,
 amsmath,amssymb,
 aps,
prb, longbibliography
]{revtex4-1}

\usepackage{graphicx}
\usepackage[danish,english]{babel}
\usepackage[utf8]{inputenc} 
\usepackage[T1]{fontenc}
\usepackage{amsmath}
\usepackage{color}

\newcommand{\LLL}{\mathbf{L}}
\newcommand{\SSS}{\mathbf{S}}
\newcommand{\rrr}{\mathbf{r}}

\newcommand{\GGG}{Gd$_3$Ga$_5$O$_{12}${}} 
\newcommand{\GAG}{Gd$_3$Al$_5$O$_{12}${}} 
\begin{document}

 \title{Spin dynamics of the director state in frustrated hyperkagome systems}
\author{Henrik Jacobsen}
\affiliation{Paul Scherrer Institute, Laboratory for Neutron Scattering and Imaging, 5232 Villigen, Switzerland} \affiliation{Department of Physics, Oxford University, Oxford, OX1 3PU, United Kingdom}
 \affiliation{Nanoscience Center, Niels Bohr Institute, University of Copenhagen, 2100 Copenhagen \O{}, Denmark}
\affiliation{European Spallation Source, Tunav\"agen 24, 223 63 Lund, Sweden}
\author{Ovidiu Florea}
\affiliation{Institut N\'eel, CNRS and Universit\'e Grenoble Alpes, 38000 Grenoble, France}
\author{Elsa Lhotel}
\affiliation{Institut N\'eel, CNRS and Universit\'e Grenoble Alpes, 38000 Grenoble, France}
\author{Kim Lefmann}
 \affiliation{Nanoscience Center, Niels Bohr Institute, University of Copenhagen, 2100 Copenhagen \O, Denmark}
\author{Oleg Petrenko}
\affiliation{Department of Physics, University of Warwick, Coventry, CV4 7AL, UK}
\author{Chris S. Knee}
\affiliation{Department of Chemical and Biological Engineering, Chalmers University of Technology, Gothenburg SE 412 96, Sweden} 
\affiliation{ESAB AB, Lindholmsallén 9, Box 8004, SE-402 77, Gothenburg, Sweden}
\author{Tilo Seydel}
\affiliation{Institute Laue-Langevin, 71 avenue des Martyrs, BP156, 38042 Grenoble Cedex 9, France}
\author{Paul F. Henry}
\affiliation{ISIS Facility, Rutherford Appleton Laboratory, Chilton, Didcot, OX11 0QX, Oxfordshire, UK}
\affiliation{Uppsala University, Department of Chemistry, \AA{}ngstr\"om Laboratory, Box 538, SE-751 21 Uppsala, Sweden}
\affiliation{Department of Chemical and Biological Engineering, Chalmers University of Technology, Gothenburg SE 412 96, Sweden}
\affiliation{European Spallation Source, Tunav\"agen 24, 223 63 Lund, Sweden}
\author{Robert Bewley}
\affiliation{ISIS Facility, Rutherford Appleton Laboratory, Chilton, Didcot, OX11 0QX, Oxfordshire, UK}
 \author{David Voneshen}
\affiliation{ISIS Facility, Rutherford Appleton Laboratory, Chilton, Didcot, OX11 0QX, Oxfordshire, UK}
\affiliation{Department of Physics, Royal Holloway University of London, Egham, TW20 0EX, UK.}
\author{Andrew Wildes}
\affiliation{Institute Laue-Langevin, 71 avenue des Martyrs, BP156, 38042 Grenoble Cedex 9, France}
\author{G\o{}ran Nilsen }
\affiliation{Institute Laue-Langevin, 71 avenue des Martyrs, BP156, 38042 Grenoble Cedex 9, France}
\affiliation{ISIS Facility, Rutherford Appleton Laboratory, Chilton, Didcot, OX11 0QX, Oxfordshire, UK}

\author{Pascale P. Deen}
\affiliation{European Spallation Source, Tunav\"agen 24, 223 63 Lund, Sweden}
\affiliation{Nanoscience Center, Niels Bohr Institute, University of Copenhagen, 2100 Copenhagen \O{}, Denmark}
 \begin{abstract}
 We present an experimental study of the magnetic structure and dynamics of two frustrated hyperkagome compounds, \GGG{} and \GAG{}. It has previously been shown that \GGG{} exhibits long-range correlations of multipolar directors, that are formed from antiferromagnetic spins on loops of ten ions. Using neutron diffraction and Reverse Monte Carlo simulations we prove the existence of similar magnetic correlations in \GAG{}, showing the ubiquity of these complex structures in frustrated hyperkagome materials.
 Using inelastic neutron scattering we shed further light on the director state and the associated low-lying magnetic excitations. In addition we have measured quasielastic dynamics that show evidence of spin diffusion. Finally, we present AC susceptibility measurements on both \GGG{} and \GAG{}, revealing a large difference in the low frequency dynamics between the two otherwise similar compounds.

 \end{abstract}

 \maketitle

\section{Introduction}
Emergence is the phenomenon of collective behavior that does not depend solely on the individual parts of a system, but rather through their interactions. This effect often results in novel and diverse states of matter. For example, in the field of magnetic frustration, emergent phenomena have been gracefully demonstrated by the observation of magnetic monopoles emerging in spin ice materials \cite{Fennell2009a, Morris2009,Castelnovo2008,Giblin2011,Dusad2019}.

\begin{figure}
 \includegraphics[width=0.35\textwidth]{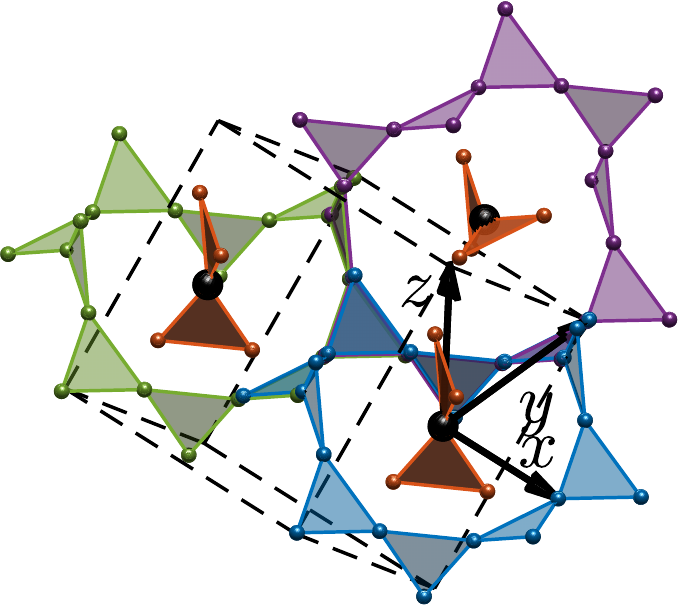}
 \hfill
 \caption{The environment of a Gd$^{3+}$ atom on the hyperkagome lattice in GGG and GAG, showing the local coordinate system and the ten-ion loop surrounding each atom. The central (black) atom is representative of the director position. The interplay between three different ten-ion loops are shown. The unit cell is shown with dashed lines. Not all Gd atoms in the unit cell are shown.
  }
 \label{fig:Fig1}
\end{figure}

A further example is the director state uncovered in the geometrically frustrated garnet, \GGG{} (GGG) \cite{Paddison2015a}. 
The  Gd$^{3+}$ ions in GGG form two interpenetrating hyperkagome lattices. In these structures, the resultant moment on connected loops of ten ions form a nematic director state  at low temperatures\cite{Paddison2015a}. Fig.~\ref{fig:Fig1} illustrates three such ten-ion loops. Taking the position of each director as the atom at the center of the ten-ion loop (illustrated in black), the directors also form a hyperkagome structure. 

The directors in GGG are reminiscent of  the emergent cluster state in the geometrically frustrated spinel ZnCr$_2$O$_4$ \cite{Lee2002}, where groups of six spins self-organise into decoupled antiferromagnetic (AF) loops, also known as director protectorates.

It has been suggested that such directors are inherent in geometrically frustrated systems, and that they provide an organising principle in which emergent clusters form out of a manifold of ground states with the low temperature dynamics governed by the director state.

In \GGG{} the director state originates from the interplay between near-neighbor AF interactions and local $xy$ anisotropy \cite{Paddison2015a}. The directors are highly anisotropic and align along their local $z$-axis. The alignment of all the directors along their local $z$ axis leads to long range correlations, in contrast to the decoupled directors in  ZnCr$_2$O$_4$. The observed magnetic excitations in GGG range from the peV to the meV scale \cite{Ghosh2008, Deen2010,Dunsiger2000,Marshall2002,Bonville2004,DAmbrumenil2015}, and the ten-ion loops seem to be central to understanding these dynamics in GGG \cite{Ghosh2008,DAmbrumenil2015}.

In this paper we delve further into the dynamics of the emergent behavior of spins on the hyperkagome lattice. We first reveal the existence of the director state in \GAG{} (GAG), isostructural to GGG, thus showing that such emergent behavior is not unique to GGG. We further probe the low energy magnetic dynamics in  GGG and GAG using AC magnetic susceptibility and inelastic neutron scattering with $\mu$eV and meV energy resolution. Finally, we discuss how the magnetic excitations can be understood in terms of a combination of single spin dynamics and collective dynamics on the ten-ion loops.

\section{Background}
GGG is a well known magnetically frustrated compound with Gd$^{3+}$ ions positioned on a three dimensional hyperkagome lattice. The Gd$^{3+}$ ions have $S=7/2$ and are usually regarded as spherically symmetric, although there are indications of a weak planar anisotropy stemming from crystal field effects \cite{Paddison2015a,Lefrancois2019}. The space group of GGG is Ia$\bar{3}$d, with $a=12.39$ \AA{}.
The Curie-Weiss temperature of GGG is $\theta_{CW}\sim-2.25$ K and neighboring spins are coupled with an antiferromagnetic exchange of $J_1=107$ mK \cite{Kinney1979}. No conventional long-range magnetic order has been found down to 25 mK \cite{Petrenko1998}. Neutron scattering revealed the onset of short-range correlations for $T < 4$ K with the director correlations developing for $T \lesssim 1.5$ K  \cite{Petrenko1998,Paddison2015a}. Below $T_f^\text{GGG}=175$ mK, the spins freeze partially \cite{Schiffer1994,Quilliam2013, Florea2017}.
The freezing is a result of the interplay between the next nearest exchange interaction, $-12$ mK$\leq J_2\leq 4$ mK, the inter-sublattice exchange interaction, $-3$ mK$<J_3<12$ mK,  and the long-range dipole interaction, $\mathcal{D}$ \cite{Yavorskii2006}. The magnetic behavior of GGG can be modeled using a standard Heisenberg model \cite{DAmbrumenil2015}:
\begin{align}
    \mathcal{H}=\sum_{ij}J_{ij}\SSS_i\cdot \SSS_j + \mathcal{D}\sum_{ij}  \frac{\SSS_i \cdot \SSS_j - 3(\SSS_i \cdot \hat{\rrr}_{ij})(\SSS_j \cdot \hat{\rrr}_{ij})}{r_{ij}^3} ,
\end{align}
where $\hat{\rrr}_{ij}$ is a unit vector between atoms $i$ and $j$. 

In comparison to GGG, GAG has a larger near-neighbor exchange constant of $J_1=186$ mK \cite{Florea2017}, and the magnitude of the Curie-Weiss constant is larger as well, $\theta_{CW}\sim -3.9(3)$ K. The further-neighbor exchange constants are presently unknown for GAG.
The $(H, T)$ phase diagram determined for GAG \cite{Florea2017} is similar to that of GGG \cite{Deen2015,Schiffer1994} with phase boundaries normalised to the magnitude of $J_1$. 
A spin freezing in GAG happens below $T_f^\text{GAG}\sim300$ mK. The dipolar interaction strength, $\mathcal{D}$, in GAG and GGG, calculated using the near-neighbor Gd distances in each compound, are similar, being slightly larger in GAG (48 mK) than in GGG (45 mK). 
The ratio $\mathcal{D}/J_1$ is thus smaller in GAG than in GGG by a factor 0.6. The parameters for GGG and GAG are summarized in Table~\ref{tab:Tab1}.

\begin{table}[htp]
    \centering
    \begin{tabular}{cccccc}
    \hline \hline
        &  $\theta_{CW}$ & $J_1$   & $\mathcal{D}$     & $\mathcal{D}/J_1$ & $T_f$ \\\hline
    GGG & $-2.25$ K      & 107 mK & 45 mK & 0.42    & 175 mK\\
    GAG & $-3.9(3)$ K      & 186 mK & 48 mK & 0.26    & 300 mK\\ \hline \hline
    \end{tabular}
    \caption{Relevant magnetic parameters for GGG and GAG as discussed in the text.}
    \label{tab:Tab1}
\end{table}

 \section{Experimental details}
We have carried out several experiments on powdered samples of GAG and GGG,  previously used in Ref.~\onlinecite{Florea2017} and Refs.~\onlinecite{Petrenko1998a,Deen2010,Deen2015} respectively. Both samples use isotopically enriched $^{160}$Gd (99.98\% for GGG, 98.4\% for GAG) to reduce the neutron absorption cross-section.

We have performed a polarized neutron powder diffraction experiment on 1.6\,g of GAG on the diffuse scattering spectrometer D7 at the Institut Laue Langevin (ILL)\cite{D7_data,D7_data2} .  The sample was cooled using a dilution refrigerator, to access the temperature regime $60$ mK$ < T < 5$~K. The magnetic signal was separated from the nuclear and spin incoherent signal using the 10-point method \cite{Ehlers2013}. We corrected for finite polarization using a quartz standard and for detector efficiency using a vanadium standard \cite{Stewart2009}. These corrections were carried out  using the Large Array Manipulation Program, LAMP \cite{Richard:1996:J.NeutronRes.}. Leaked signals at $Q$-values corresponding to strong nuclear Bragg peaks from the copper sample container  were removed. The data were normalized to mbarns/sr/Gd$^{3+}$ by refining the nuclear structure using Fullprof \cite{Rodriguez-Carvajal1993}.

The magnetic dynamics of powdered GAG on the meV scale were measured using the cold neutron multi-chopper spectrometer LET at ISIS \cite{Bewley2011,LET_data} using incident energies $E_{i}$ = 1.25, 2.0, 3.96 and 8.69 meV with an elastic energy resolution of 0.023(1), 0.042(1), 0.11(1)  and 0.42(1) meV, respectively (full width at half maximum (FWHM)), determined using an incoherent scatterer. The data were reduced using Mantid \cite{Arnold2014}. We measured the magnetic dynamics at three temperatures (50~mK, 500~mK and 1~K). 
The slow magnetic dynamics of powdered GGG on the $\mu$eV scale were measured using the neutron backscattering instrument IN16b of the ILL \cite{IN16b_data,Frick2010}, across a temperature range $60$ mK$ < T < 1$ K, using a dilution refrigerator. The neutron data were converted from acquisition channel to energy using standard treatments using LAMP \cite{Richard:1996:J.NeutronRes.}.  The instrumental resolution function can be described by a Gaussian with weak Lorentzian tails, FWHM of 0.94(2) $\mu$eV, determined using a vanadium standard.

In all neutron scattering experiments the samples were cooled for 24 hours prior to measurements to ensure that base temperature was reached.

AC susceptibility measurements were performed using a superconducting quantum interference device magnetometer equipped with a miniature dilution refrigerator, developed at the Institut N\'{e}el, Grenoble \cite{ExtractionMethod}. Powders of GGG (17.13 mg) and GAG (5.7 mg) were cooled in zero applied magnetic field, and measured at temperatures down to 80 mK with the frequency varied between 0.057 Hz and 1110 Hz under an applied field of $\mu_{0}H_{AC}$ = 0.055 mT. In these low temperature measurements, the samples were mixed with Apiezon N grease to ensure thermalization. Furthermore, a single crystal of GGG was measured at frequencies down to 1 mHz with the AC field applied along the (110) direction. 
 \section{Results and discussion}
 
\subsection{Magnetic structure of GAG}

\begin{figure}[t]
    \centering
    \includegraphics[width=0.44\textwidth]{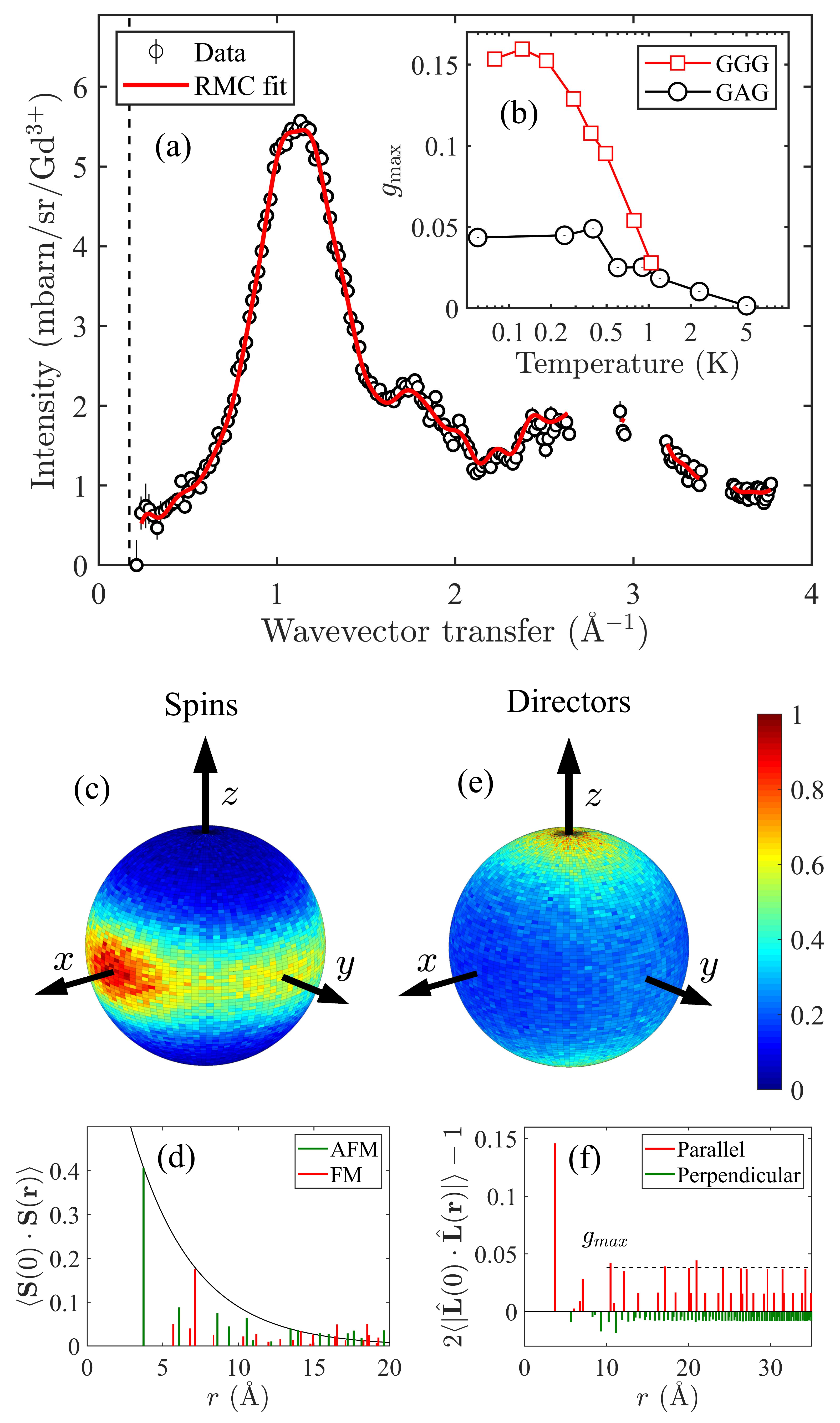}
    \caption{ (a) Magnetic $S(Q)$ at $T=60$\,mK for GAG. The red line is a RMC fit using SPINVERT as described in the text. The dashed line indicates the lower limit for the RMC calculation. (b) Temperature dependence of the loop order parameter $g_\text{max}$ for  GAG and GGG \cite{Paddison2015a}. 
    (c) The relative distribution of spin orientations at $T=60$\,mK in the local coordinate system shown in Fig.~\ref{fig:Fig1}. 
    (d) The average spin-spin correlation at $T=60$ mK as a function of the distance between the spins, $r$. Green bars show antiferromagnetic correlations, $\langle \SSS(0) \cdot \SSS(\rrr)\rangle<0$, while red bars show ferromagnetic correlations,  $\langle \SSS(0) \cdot \SSS(\rrr)\rangle>0$. The black line is the function $\exp(-r/\xi)$ with $\xi=4.2(1)$\,\AA{}.   
    (e) The relative distribution of director orientations at $T=60$\,mK in the local coordinate system shown in Fig.~\ref{fig:Fig1}.
    (f) The correlation, $g=2\langle | \hat{\LLL}(0) \cdot \hat{\LLL}(\rrr) | \rangle -1$, between directors at $T=60$ mK as a function of the distance between them $r$. The dashed line indicates the saturation level,  $g_\text{max}$, at high $r$.
}
\label{fig:Fig2}
\end{figure}

Fig.~\ref{fig:Fig2}(a) shows the magnetic scattering cross section, $S({Q})$, of powdered GAG as a function of wave vector transfer, $Q$, at 60~mK, measured at D7. It is worth noting that no energy analysis is performed on D7, and the observed signal is thus integrated over energy. This means that low energy magnetic excitations contribute to the observed diffraction signal. However, from the LET experiment on GAG presented in the next section we find that 97.4\% of the energy integrated signal is within the elastic line, and the low energy magnetic excitations thus contribute only 2.6\% to the total magnetic and nuclear scattering. At the peak near $Q=1.1$ \AA{}$^{-1}$ more than half of the total signal observed at D7 is magnetic. Therefore, the observed magnetic signal at D7 is at least $\sim95$\% elastic scattering, and the small contribution from inelastic scattering can safely be ignored. 

The broad diffuse peaks around $Q=1.1$~\AA{}$^{-1}$, $Q=1.8$~\AA{}$^{-1}$ and $Q=2.9$~\AA{}$^{-1}$ are similar to those observed in GGG \cite{Petrenko1998}, and indicate the presence of short-range magnetic order. In GGG, a detailed Reverse Monte Carlo (RMC) analysis further showed that the shape of the peaks at $Q=1.1$~\AA{}$^{-1}$ and $1.8$~\AA{}$^{-1}$, is a signature of the director correlations \cite{Paddison2015a}. We have carried out a similar analysis of our GAG data using the RMC program SPINVERT as explained below \cite{Paddison2013a}.

The SPINVERT analysis was performed using a magnetic supercell of $6\times 6\times 6$ unit cells with a spin on each of the magnetic sites. We found $n=6$ (5184 spins) to be the smallest unit cell to yield a refinement consistent with our data. The spins were treated as classical vectors, an approximation which is justified by the large spin of Gd$^{3+}$ and the localized nature of the spins. Each simulation was initiated with every spin pointing along a random direction. In each Monte Carlo step a random spin was rotated slightly, and the powder averaged magnetic neutron diffraction signal from the spin configuration was calculated and scaled by a constant scale factor, $s=g^2S(S+1)=63$, then compared with the experimental data. The new spin orientation was kept or rejected using the Metropolis algorithm \cite{Metropolis1949}. We carried out 100 independent simulations for each temperature, with 100 steps per Gd$^{3+}$ ion, resulting in 518400 steps for each simulation. Data presented are an average of these 100 simulations.  We note that unlike direct Monte Carlo, this procedure provides no information on the relevant exchange interactions, but does yield a spin structure consistent with the neutron scattering data.

The resulting RMC fit is shown as a continuous red line through the data points in Fig.~\ref{fig:Fig2}(a), showing excellent agreement with the data. The finite system size makes the calculation unreliable for $Q< 2\pi/(na/2) \leq 0.17$~\AA{}$^{-1}$, indicated by the dashed black line. 

We now look at the RMC spin structure to determine the local spin directions and any director correlations. Fig.~\ref{fig:Fig2}(c) shows a three dimensional histogram of the local orientation of each Gd$^{3+}$ spin at 60 mK, normalized by the solid angle of the bin. Similar to GGG we find a strong local $xy$-anisotropy. It is possible that the anisotropy originates from the dipolar interaction \cite{Paddison2015a}.   
Fig.~\ref{fig:Fig2}(d) shows the magnitude of the spin-spin correlations averaged over shells at distance $r$,  $|\langle \SSS(0)\cdot \SSS(\rrr)\rangle|$. We find antiferromagnetic near-neighbor correlations which follow an approximate exponentially decay, $\langle \SSS(0) \cdot \SSS(\rrr) \rangle \approx  \exp(-r/\xi)$, with $\xi= 4.2(1)$\,\AA{}. For near neighbors we find $\langle \SSS(0) \cdot \SSS(\rrr)\rangle=-0.41(1)$. This corresponds to an average angle between near-neighbor spins of 114$^{\circ}$, significantly lower than the 120$^{\circ}$ expected for a pure Heisenberg spin state. In GGG a much smaller deviation was found, 118$^{\circ}$.

In GGG it was found that the spin structure depends strongly on the precise values of $J_1$, $J_2$ and $J_3$, and their interplay with the dipolar interaction \cite{Yavorskii2006}. The same is true for GAG, but since $J_2$ and $J_3$ presently are unknown in GAG, we cannot quantify their effect on the spin structure. 

In order to look for evidence of the director state we follow Ref.~\onlinecite{Paddison2015a} and define the 10-spin directors, $\LLL$ as the staggered magnetization of a loop
\begin{align}
 \LLL = \sum_{n=1}^{10} (-1)^n \SSS_n.
\end{align}

The distribution of directors in GAG is shown in Fig.~\ref{fig:Fig2}(e), revealing a strong local $z$ anisotropy, as also found in GGG. The correlation between directors can be quantified as
\begin{align}
 g(r)=2\langle | \hat{\LLL}(0) \cdot \hat{\LLL}(\rrr) | \rangle -1,
\end{align}
where $\hat{\LLL}=\LLL/L$. The director-director correlation is $g=1$ for directors that are parallel, and $g=-1$ for directors that are orthogonal. We show $g(r)$ in Fig.~\ref{fig:Fig2}(f). The near-neighbor director-director correlation is strong, which is unsurprising, as neighboring directors share several spins. Importantly, the director-director correlations then falls to a constant value, $g_\text{max}$, that extends beyond 30 \AA{}. As such we find that the directors in GAG remain correlated, albeit weakly, over long distances.  

Fig. \ref{fig:Fig2}(b) shows the temperature dependence of $g_\text{max}$ for  GGG and GAG. Both compounds show an increase in $g_\text{max}$  below 1.5 K, as the director state emerges. $g_\text{max}$ increases rapidly with decreasing temperatures, and peaks at $\sim$400 mK and 175 mK for GAG and GGG, respectively. 
Below this temperature, which coincides roughly with the onset of the spin freezing \cite{Schiffer1994,Quilliam2013, Florea2017},  $g_\text{max}$ appears to remain constant.
$g_\text{max}(\text{GGG}) \approx 3g_\text{max}(\text{GAG})$ indicating that the director-director correlations at the lowest temperatures are stronger in GGG than in GAG. 

This difference is likely caused by a combination of variations in $J_2$ and $J_3$, as well as the variation in $\mathcal{D}/J_1$, with $\mathcal{D}/J_1$ for GAG being smaller by a factor 0.6. It would be interesting to study other Gd$^{3+}$ based garnets with different values of $\mathcal{D}/J$. For example, Gd$_3$Te$_2$Li$_3$O$_{12}$ (GTLG) has $\mathcal{D}/J=0.36$\cite{Quilliam2013}, and we therefore predict a director state to be present in GTLG with $0.05<g_\text{max}^\text{GTLG}<0.15$. Ref.~\cite{Mukherjee2017} also suggests multiple Gd$^{3+}$-based garnets with varying values of $\mathcal{D}/J_1$ that would be interesting to study. It would also be valuable to determine $J_2$ and $J_3$ for GAG. For GGG, the most accurate determination of these parameters was based on an analysis of the longer-range correlations that set in in GGG for temperatures below 175 mK \cite{Yavorskii2006}. Such correlations are not present in GAG, making a similar analysis impossible at present. 

The presence of director correlations in GAG as well as in GGG indicate that any hyperkagome system with antiferromagnetic near-neighbor interactions and relatively strong dipolar interactions should display a director state. 

A recent study of the isostructural compound 
Yb$_3$Ga$_5$O$_{12}$, (YbGG), also reveals a director state. However, in stark contrast to GGG and GAG,  the director state in YbGG is derived from ferromagnetic near-neighbor exchange, with spins directed along the local $z$ direction \cite{Sandberg2020}, with directors correlated over short distances. Despite these differences, the director state on hyperkagome lattices thus appears to be rather robust. To form a director state, some form of anisotropy in the spin orientation seems to be required. In GGG and GAG, this anisotropy is most likely a result of the dipolar interaction, with the potential addition of small crystal field effects, whereas the anisotropy in YbGG originates almost entirely from crystal field effects,. The precise interactions which determine whether the director state is long- or short-range are unclear. 

\subsection{meV dynamics of GAG}
\begin{figure}[htb]
 \includegraphics[width=0.48\textwidth]{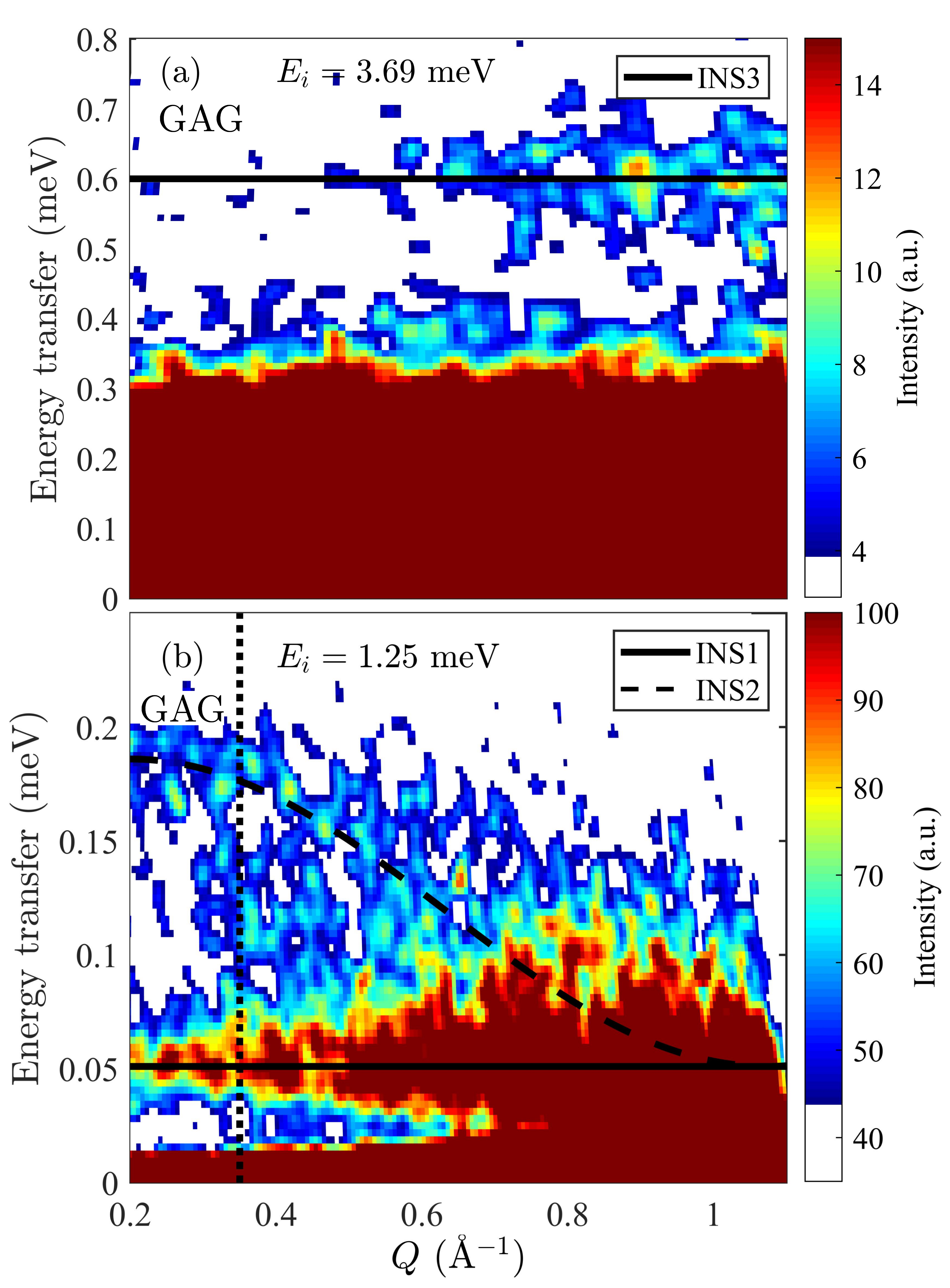}
 \caption{$S(Q, E$) of GAG at 50 mK measured at LET with incident neutron energy  (a) $E_{i}$ = 3.69  and (b) 1.25 meV. The dashed line on (b) indicates the central position of the cuts used to quantify the temperature variations of these excitations in Figs.~\ref{fig:Fig4} and \ref{fig:Fig5}. The lines are guides to the eye, and the color scale is chosen to emphasize INS1, INS2 and INS3.}
  \label{fig:Fig3}
 \end{figure}

Fig.~\ref{fig:Fig3} shows $S(Q,E)$, at 50~mK for GAG measured at LET using incident energies $E_{i}$ = 3.69 and 2.0 meV. We observe three distinct features, similar to previous results on GGG \cite{Deen2010}. We label the three features INS1, INS2 and INS3 in line with the nomenclature employed for GGG.

INS3 is a weak non-dispersive excitation observed at $0.588(5)$~meV, shown in Fig.~\ref{fig:Fig3}(a). The intensity of this excitation does not follow a single ion form factor, and is thus not a crystal field excitation. In GGG this excitation was found at 0.58(3) meV \cite{Deen2010}. At lower energies, Fig.~\ref{fig:Fig3}(b), we observe two further distinct features, labelled INS1 and INS2.  Within the energy resolution and statistics afforded by the experiment INS1 is a low-lying dispersionless excitation at $E = 0.05(1)$~meV with INS2, a much broader feature that is more dispersive in nature. At low $Q$, INS2 appears at $E=0.19$ meV and falls into the elastic line near $Q\sim1$ \AA{}$^{-1}$. All three excitations have slightly higher energies in GAG than in GGG, consistent with the stronger exchange interactions in GAG.

\begin{figure}[tb]
  \includegraphics[width=0.48\textwidth]{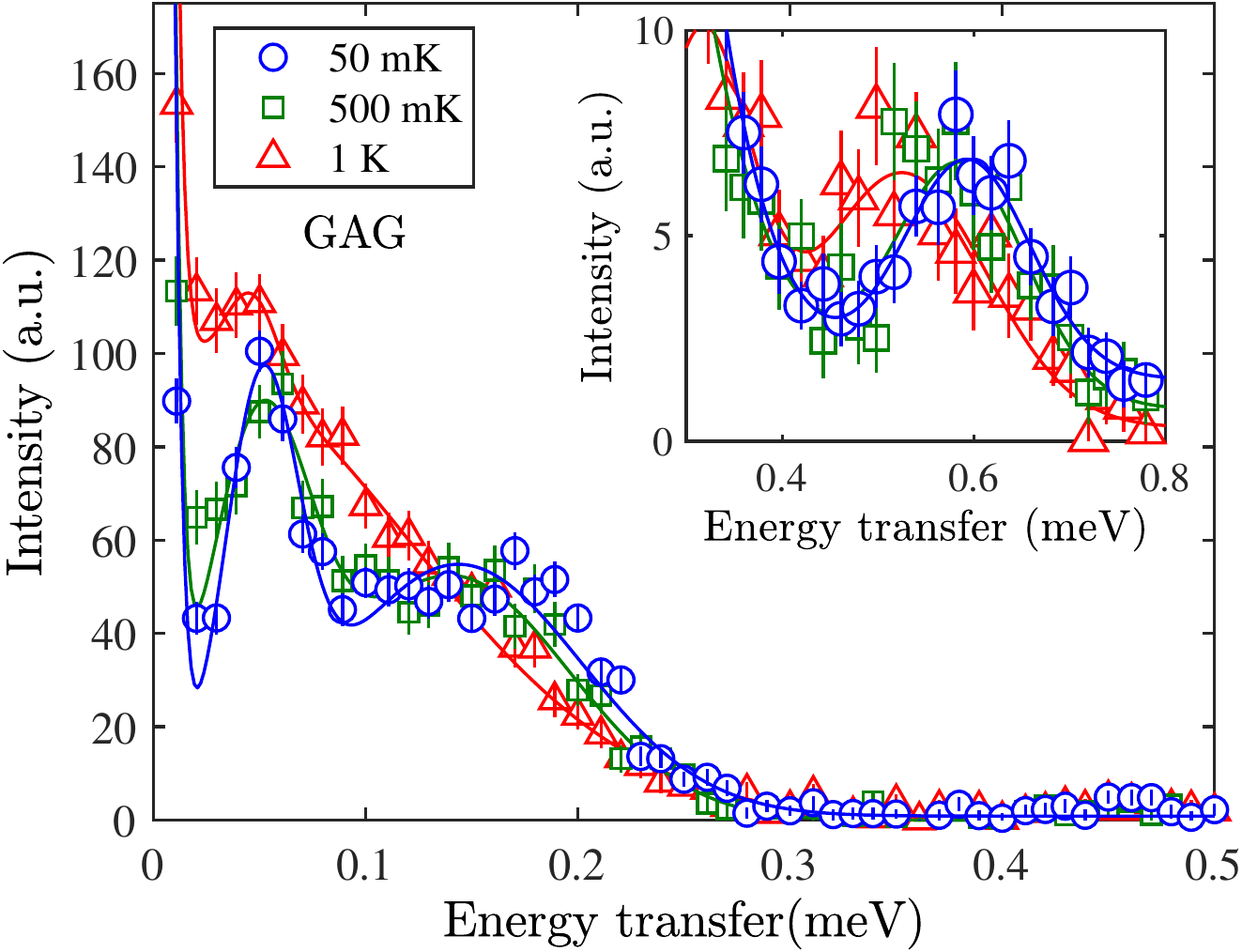}
  \caption{Temperature and energy dependence of INS1 and INS2 for GAG, measured at LET. A 0.2~\AA{}$^{-1}$ wide cut of the data around $Q=0.35$~\AA{}$^{-1}$ at 0.05, 0.5 and 1 K, shows the existence of two low-lying excitations. The inset shows the $Q$ integrated intensity of INS3. Continuous lines are fits to the data. 
  }
  \label{fig:Fig4}
 \end{figure}

The dotted vertical line in  Fig.~\ref{fig:Fig3}(b) shows the direction of cuts through the data for a range of temperatures $T = 50$ mK, 500 mK and 1 K.  The cuts are presented in Fig.~\ref{fig:Fig4}(b).
At 50 mK the three excitations are well defined and can be modeled as gapped modes. Upon increasing the temperature, the signal broadens. The gap to INS1 remains visible, but INS2 appears to become overdamped and thus transforms into almost featureless quasi-elastic scattering. INS3, in contrast, remains well defined at least up to 1 K. 
 
\begin{figure}[htb]
  \includegraphics[width=0.48\textwidth]{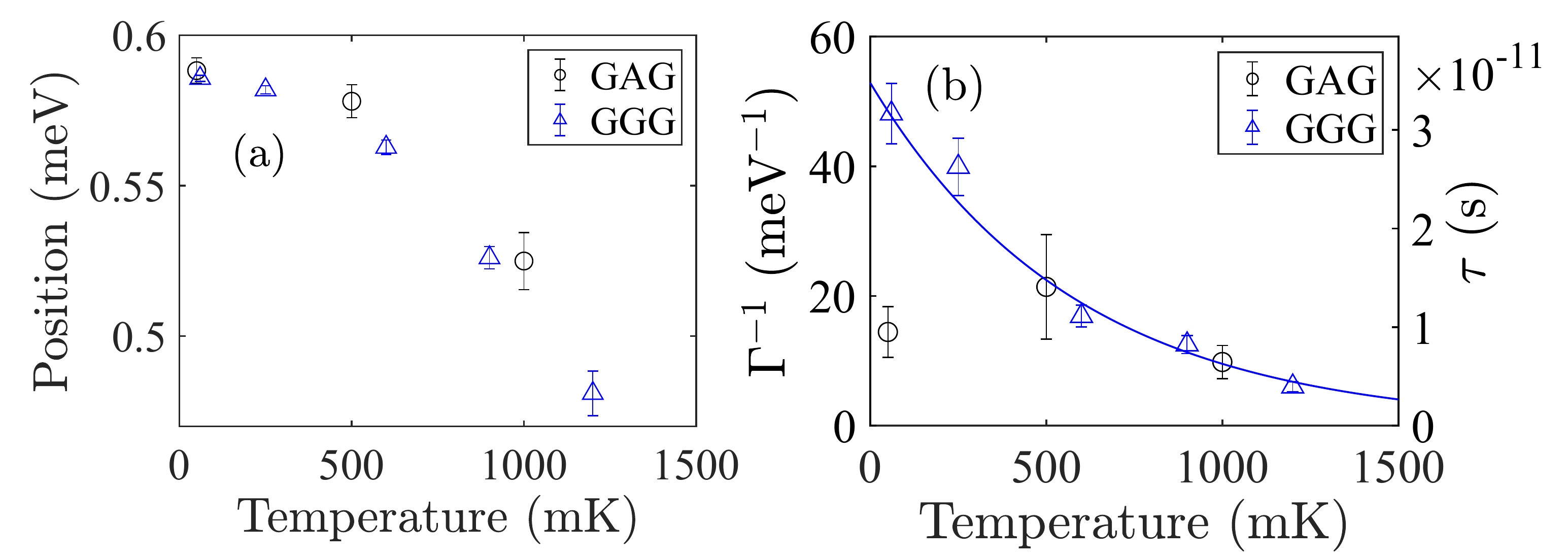}
\caption{(a) Temperature dependence of the position of the INS3 excitation for GAG and a GGG. (b) Temperature dependence of the fluctuation rate expressed as the inverse of the HWHM of the peak. The solid line is a guide to the eye.   }
  \label{fig:Fig5}
 \end{figure}

The temperature dependence of the position of INS3 is similar to GGG, see Fig.~\ref{fig:Fig5}(a). Fig.~\ref{fig:Fig5}(b) shows the resolution-corrected half width at half max ($\Gamma$) of the INS3 peak in GAG and GGG and the excitation life-time, $\tau=\hbar/\Gamma$. At elevated temperatures the excitation life time is nearly identical in the two compounds \cite{Deen2010}. The solid line is a guide to the eye. 

A magnetic field study by d'Ambrumenil et al.  of the magnetic excitations in GGG have previously provided theoretical insight into the excitations of GGG \cite{DAmbrumenil2015}. d'Ambrumenil et al. accurately reproduced the full dispersion of GGG in an applied magnetic field using linear spin-wave theory on the ten-ion loops, including near-neighbor interactions and the dipolar interaction. This study was carried out in a magnetic field large enough to create a ferromagnetic ground state, very different from the zero-field ground state. Still, the dispersion is similar in both applied and zero field, indicating that the underlying excitations are similar. Comparing our results with those of Ref.~\cite{DAmbrumenil2015} suggests that INS1 originates from bands that are near dispersion-less across the entire Brillouin Zone. INS2, on the other hand, consists of multiple dispersive excitations that cross and overlap in reciprocal space. 

No equivalent of INS3 is found by linear spin wave theory in the ordered state, and it may therefore have an entirely different origin. The mode is relatively sharp in energy, and has a strikingly similar position and width in the two compounds. The lack of scaling with $J_1$ suggests that near-neighbor interactions is not the main driver of this excitation. In GGG, the INS3 mode was modeled as a singlet-triplet excitation, based on the $Q$-dependence of the intensity \cite{Deen2010}. Due to the weak signal we could not extract the $Q$-dependence of the intensity in GAG, and thus cannot verify if this model fits the present data. 

\subsection{$\mu$eV dynamics of GGG} \label{sec:IN16b}
  \begin{figure}[ht]
  \includegraphics[width=0.35\textwidth]{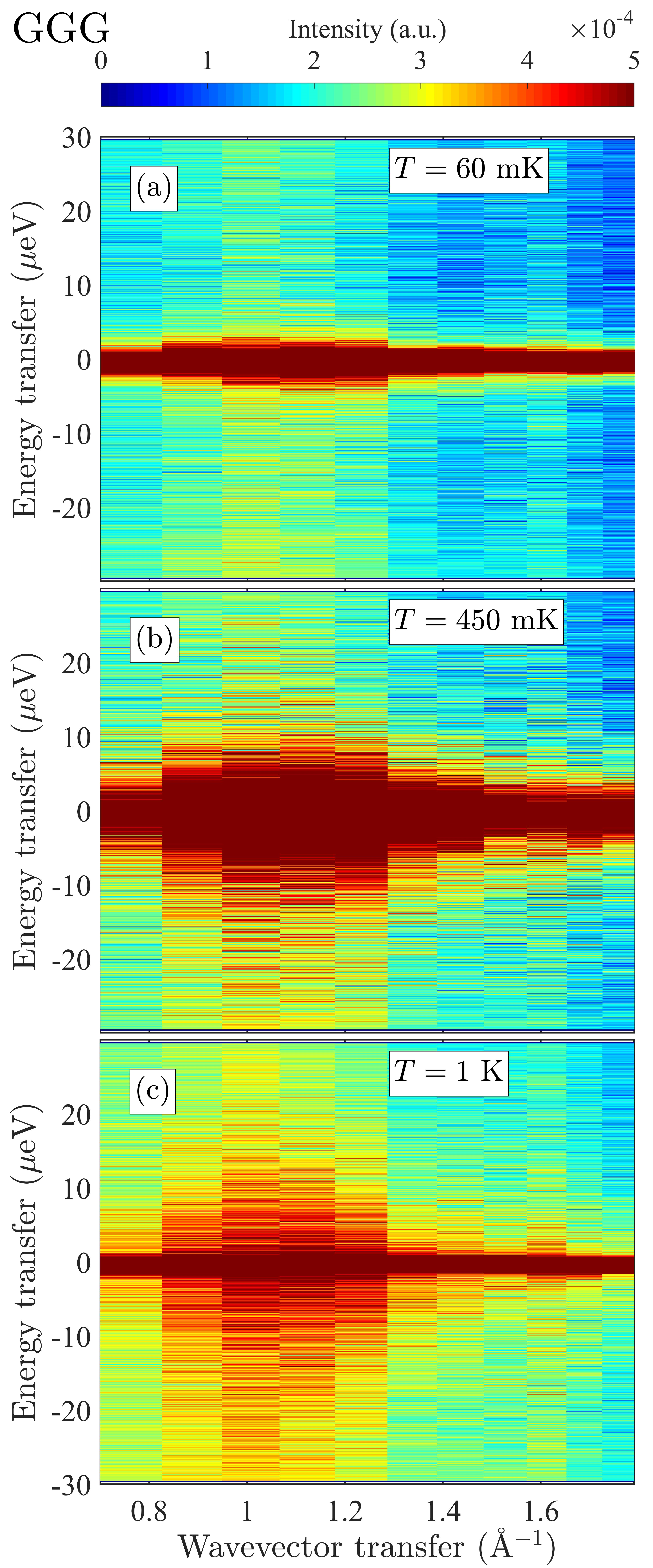}
\caption{$S(Q,E)$  of GGG measured at IN16b at temperatures $T =60$ mK, 450 mK and 1 K. 
}
  \label{fig:Fig6}
 \end{figure}

An overview of the backscattering data, $S(Q, E)$, for GGG is shown in Fig. \ref{fig:Fig6} for $60$ mK$ < T < 1$ K. At all temperatures a strong elastic signal is present for all accessible wavevector transfers. At the lowest temperature probed, 60 mK, the signal appears almost entirely elastic. A broad quasielastic broadening is observed around 450 mK with a peak in intensity near wavevevector transfer $Q=1.1$~\AA{}$^{-1}$. At 1 K the quasi-elastic signal has broadened further and reaches the edge of the experimental energy window. 
 \begin{figure}[htp]
 \includegraphics[width=0.48\textwidth]{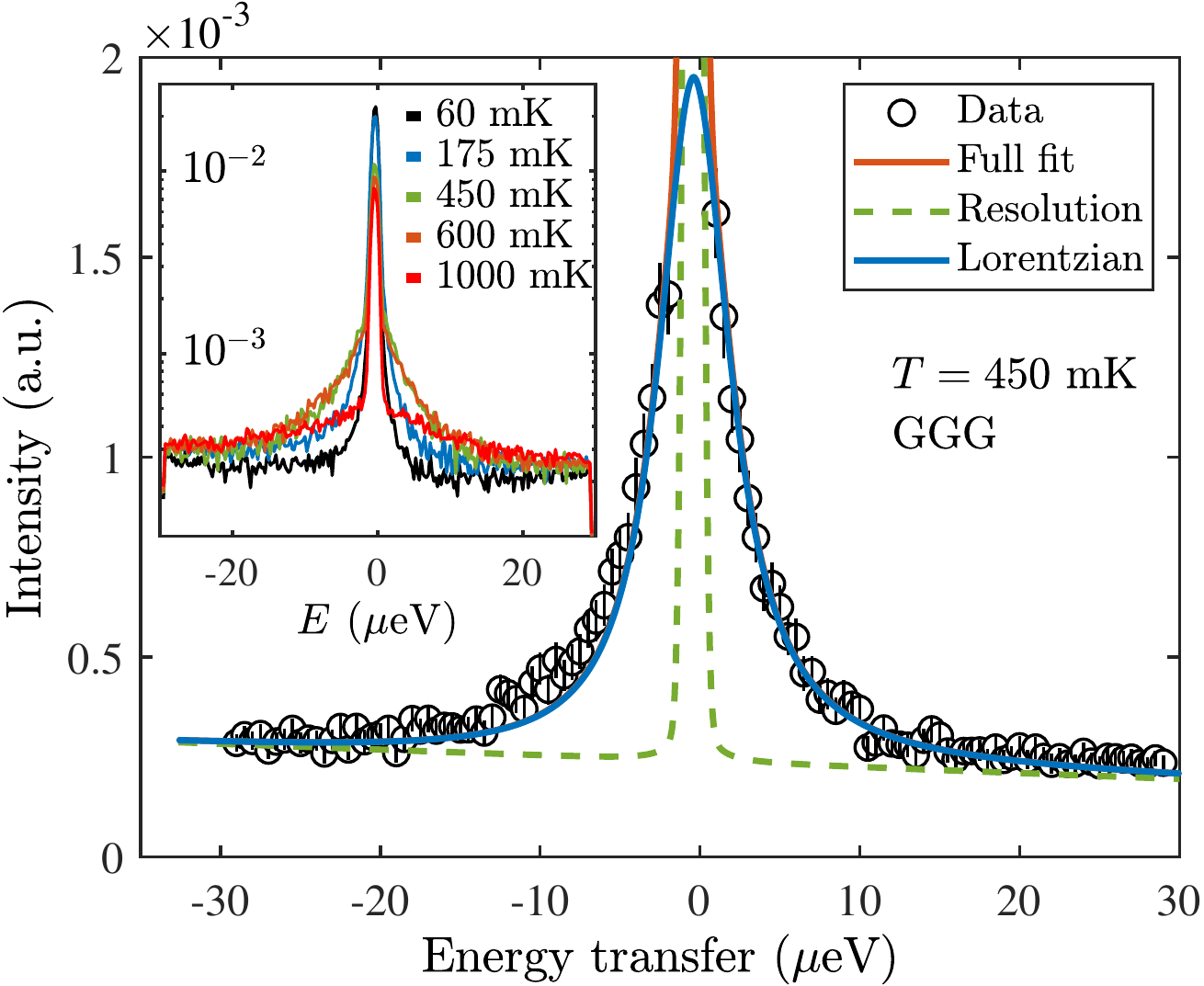}
 \caption{Example of IN16b data on GGG at 450 mK for $Q=1.1$ \AA{}$^{-1}$. The solid red line is a fit as described in the text. The solid blue line is the Lorentzian contribution. The dashed green line shows the resolution function on top of the background. The inset shows the temperature dependence of the data on a log scale.}
 \label{fig:Fig7}
\end{figure}

 The energy dependence of the total scattering signal is well described, for all scattering vectors and temperatures, by a convolution of the resolution function with a temperature independent elastic term, represented by a delta function, and a temperature dependent quasi-elastic Lorentzian term
 \begin{align}
     I=R\otimes \left(A_\text{el} \delta(E)+\frac{A_\text{qel} }{\pi} \frac{\Gamma}{\Gamma^2+E^2}  \right) + C_1+C_2E,
 \end{align}
 where $R$ is the resolution function, $A_\text{el}$ is the integrated intensity of the elastic signal, and $A_\text{qel}$ is the intensity of the quasielastic signal, $2\Gamma$ is the FWHM of the quasielastic signal. A linear sloping background is included, described by the parameters $C_1$ and $C_2$. We have accounted for detailed balancing in the fitting \cite{Boothroyd2020}, although the effect is minimal.  Fig.~\ref{fig:Fig7} shows an example of the data (black circles) and fit (red solid line) at 450 mK and $Q=1.1$~\AA{}$^{-1}$. We see a significant quasielastic broadening, compared with the instrumental resolution (dashed green line). 
 
 The inset of Fig.~\ref{fig:Fig7} shows the temperature dependence of the wavevector integrated intensity, which follows the trend seen in Fig.~\ref{fig:Fig6}. The temperature dependence of the Lorentzian signal indicates its magnetic origin. 
 
 The most important fitted parameters are shown in Fig.~\ref{fig:Fig8}, and the background parameters are discussed in the appendix.  The $Q$- and temperature dependence of the integrated intensity of the Lorentzian signal, $A_\text{qel}$, is shown in Fig.~\ref{fig:Fig8}(a). The integrated intensity, with broad peaks at $Q$=1.1 and 1.8 \AA{}$^{-1}$, follows the $Q$ dependence of the elastic magnetic structure factor previously determined for GGG\cite{Petrenko1998a}, and shown for GAG in Fig.~\ref{fig:Fig2}(a).  
The temperature dependence of the integrated intensity of the quasi-elastic component, $A_\text{qel}$ at $Q=1.1$~\AA{}$^{-1}$ is shown in Fig.~\ref{fig:Fig8}(c). It remains roughly constant for $T < 0.2$ K and decreases rapidly with increasing temperature, consistent with the decrease of magnetic correlations with increasing temperature. 

The temperature and $Q$ dependence of the FWHM of the  quasi-elastic signal, $2\Gamma$, is shown in Fig.~\ref{fig:Fig8}(b). It is interesting to note that $\Gamma$ is independent of $Q$ in the accessible range ($0.7$\,\AA$^{-1}<Q<1.8$\,\AA$^{-1}$), but increases rapidly with increasing temperature. The lack of $Q$-dependence indicates that the spin fluctuations are uncorrelated on this energy scale, a surprising result given the existence of both director and static spin-spin correlations in GGG. 

\begin{figure}
\includegraphics[width=0.5\textwidth]{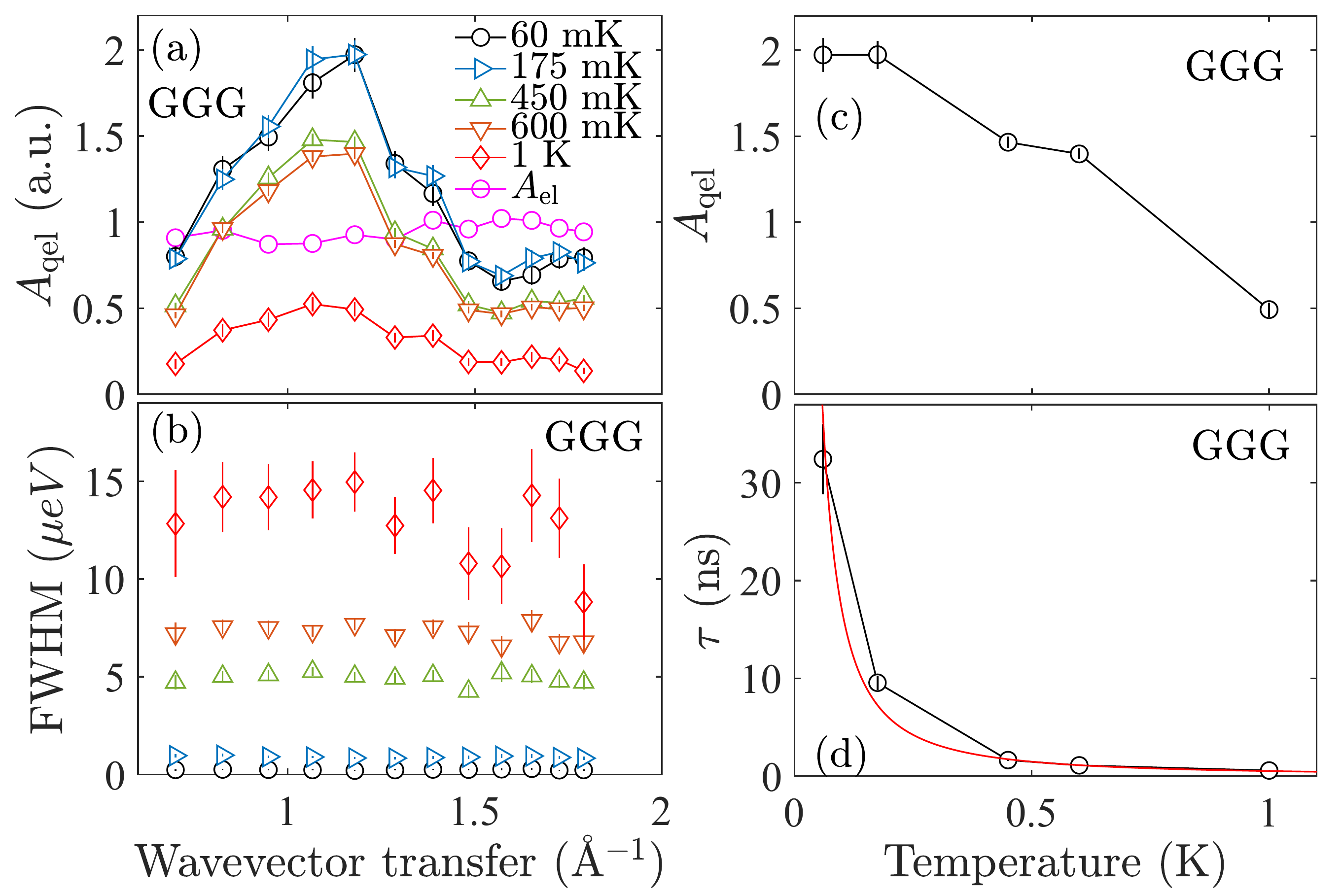}
 \caption{(a) $Q$ dependence of the area of the Lorentzian signal, $A_\text{qel}$ measured on GGG at IN16b, showing peaks at $Q$=1.1 and 1.8 \AA{}$^{-1}$, indicating that this signal is magnetic. (b) The resolution-corrected FWHM of the Lorentzian signal ($2\Gamma$), displaying no significant $Q$ dependence. 
 (c) Temperature dependence of $A_\text{qel}$. (d) Temperature dependence of the relaxation rate, $\tau=\hbar/\Gamma$, along with a fit to $\tau=AT^{B}$, with $B=-1.5(3)$. 
 }
 \label{fig:Fig8}
\end{figure}

A single Lorentzian lineshape indicates that the temporal spin-spin correlations decay as
\begin{align}
 \langle {\bf S}(0) \cdot {\bf S(t)}\rangle \sim e^{-t/\tau},
\end{align}
where the decay rate, $\tau$ is related to the FWHM by $2\Gamma=2\hbar/\tau$ \cite{Mirebeau2008}.
The temperature dependence of $\tau$ at  $Q=1.1$\,\AA{}$^{-1}$ is shown in Fig. \ref{fig:Fig8}(d) and follows $\tau=A T^B$, with $B=-1.5(3)$. 
Our results are consistent with earlier M\"ossbauer measurements on GGG \cite{Bonville2004} as well as muon spin rotation measurements \cite{Marshall2002,Dunsiger2000}, who all found spin fluctuations in GGG on a similar time scale as presented here. In the M\"ossbauer measurements, the temperature dependence was quadratic \cite{Bonville2004} and the fluctuations were shown to be confined to a plane, consistent with the spin anisotropy mentioned earlier \cite{Paddison2015a}. 

Significant theoretical work has been performed on the spin fluctuation rate for a system of classical spins on frustrated lattices. Conlon and Chalker predicted, for Heisenberg spins on a pyrochlore lattice in a cooperative paramagnetic regime,  a linear temperature dependence of the spin fluctuation rate ($\tau^{-1}$) with spin correlations that relax at a rate independent of the wavevector transfer  \cite{Conlon2009}  The hyperkagome lattice can be viewed as a depleted pyrochlore lattice, where each tetrahedron is missing a site. As such, we can expect the magnetic behavior on these two systems to be similar. 
In GGG, for $T < 1$ K, the spin correlations are indeed independent of $Q$, although $\tau$ follows a power law rather than being linear. 

More recently, the diffusive dynamics on frustrated kagome lattices have been investigated. Taillefumier predicts a distribution of timescales in the cooperative paramagnetic regime representing loops diffusing in an entropically dominated free energy landscape. The signature of such diffusion can be observed for $Q <  2\pi/d$ where $d = a$ is the loop spatial scale. In this $Q$ region, $\tau^{-1}$  increases linearly with $Q$ \cite{Taillefumier2014}. Unfortunately, we do not access a wavevector transfer below 0.7 \AA{}$^{-1}$, and thus do not directly observe this. However, in line with Taillefumier, at larger wavevector transfers the fluctuation rate is proportional to temperature and independent of $Q$ \cite{Taillefumier2014}. We therefore suggest that we are probing the diffusive dynamics of the multipolar director state in GGG. Data at lower $Q$ is required to confirm or reject this hypothesis. Such experiments will be challenging due to the vanishing intensity at low $Q$ as seen in diffraction experiments.

Further progress on understanding the director dynamics on the $\mu$eV and meV energy scale will also require inelastic neutron scattering experiments on single crystal samples. Such experiments are only feasible on isotope enriched single crystals to minimize the absorption of cold neutrons required to achieve the relevant energy resolutions. The cost of such crystals is prohibitively high. 
 
Finally, we note that some magnetic scattering is seen in the apparent background signal, as it depends on temperature (see the appendix for details). This indicates that part of the background originates from motions on a larger energy scale than the accessible energy window of IN16b ($\sim \pm 30$~$\mu$eV), as also found in similar backscattering measurements on Tb$_2$Sn$_2$O$_7$ \cite{Mirebeau2008}. It is likely that this contribution originates from the tails of the lowest excitation, INS1, observed at 0.05(1) meV at the lowest temperature.

\subsection{AC susceptibility}
\label{sec:GGG_susceptibility}
 AC susceptibility is a very sensitive probe of low frequency magnetic dynamics \cite{Topping2019}. In our AC susceptibility measurements we measure the real ($\chi'$) and imaginary susceptibility ($\chi''$), for both GGG and GAG as a function of temperature. 

\begin{figure}[ht]
\includegraphics[width=0.48\textwidth]{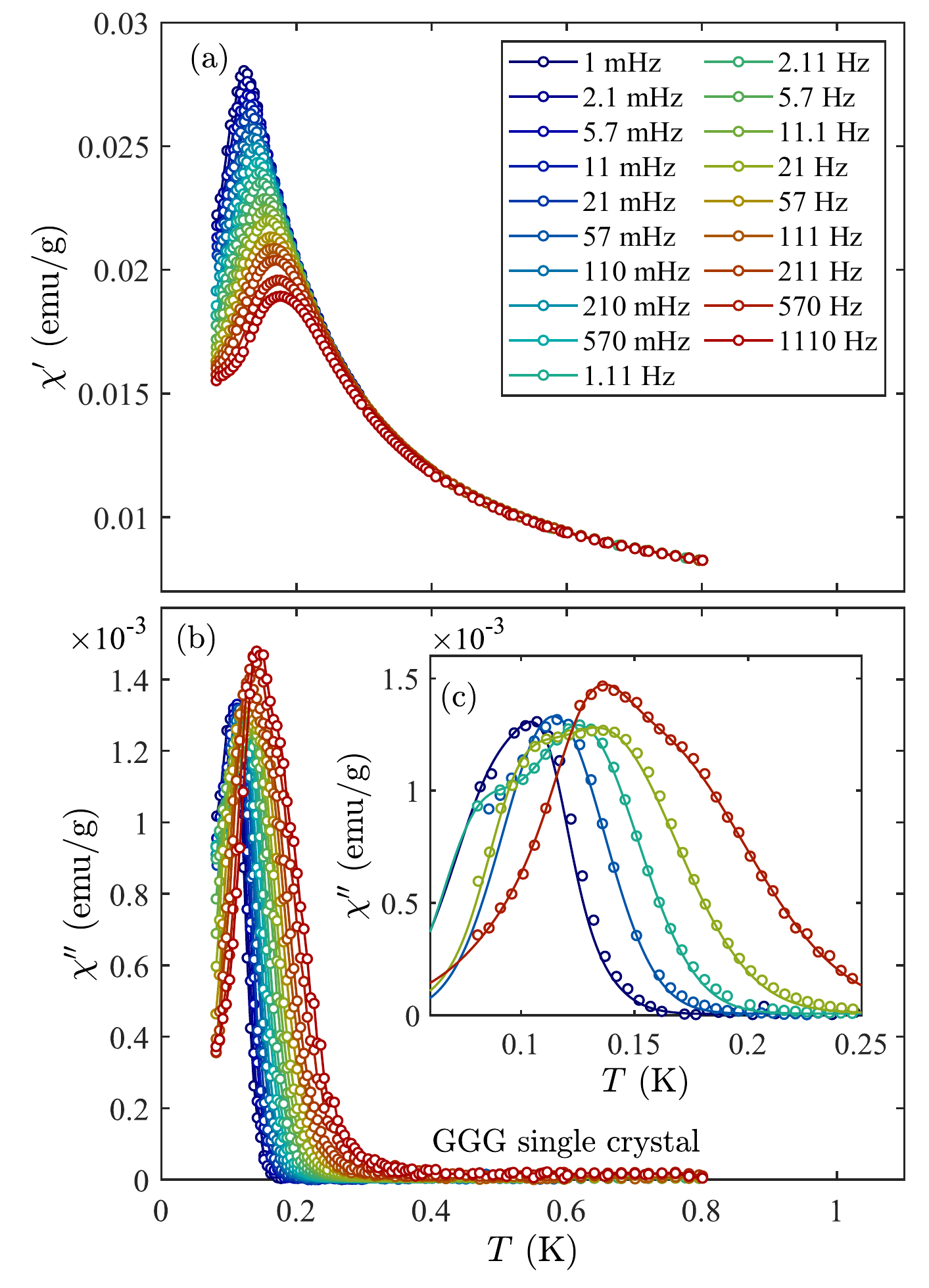}
 \caption{
 Temperature and frequency dependence of (a) the real ($\chi'$ and (b) imaginary ($\chi''$) AC susceptibliity on a single crystal of GGG measured along the (110) direction. The frequencies, $1$ mHz$ < f < 1110$ Hz, are indicated by color.
 (c) Detailed view of $\chi''$ for selected frequencies. Continuous lines are fits to Gaussians as described in the text.
} 
 \label{fig:Fig9}
\end{figure}

Fig. ~\ref{fig:Fig9}(a) shows $\chi'$ for a single crystal of GGG with the AC field applied along the $[110]$ direction. An anomaly around $T_f\sim175$ mK is seen, with a slight frequency-dependence.
Fig. ~\ref{fig:Fig9}(b) shows $\chi''$ for the same single crystal, with (c) zooming in on selected frequencies. The temperature dependence of $\chi''$ varies from a single peak for frequencies below $\sim 5$~Hz to a double peak structure emerging for frequencies $>$ 5 Hz. We fitted the peaks using Gaussian line shapes to extract the position of the peaks.

\begin{figure}
\includegraphics[width=0.38\textwidth]{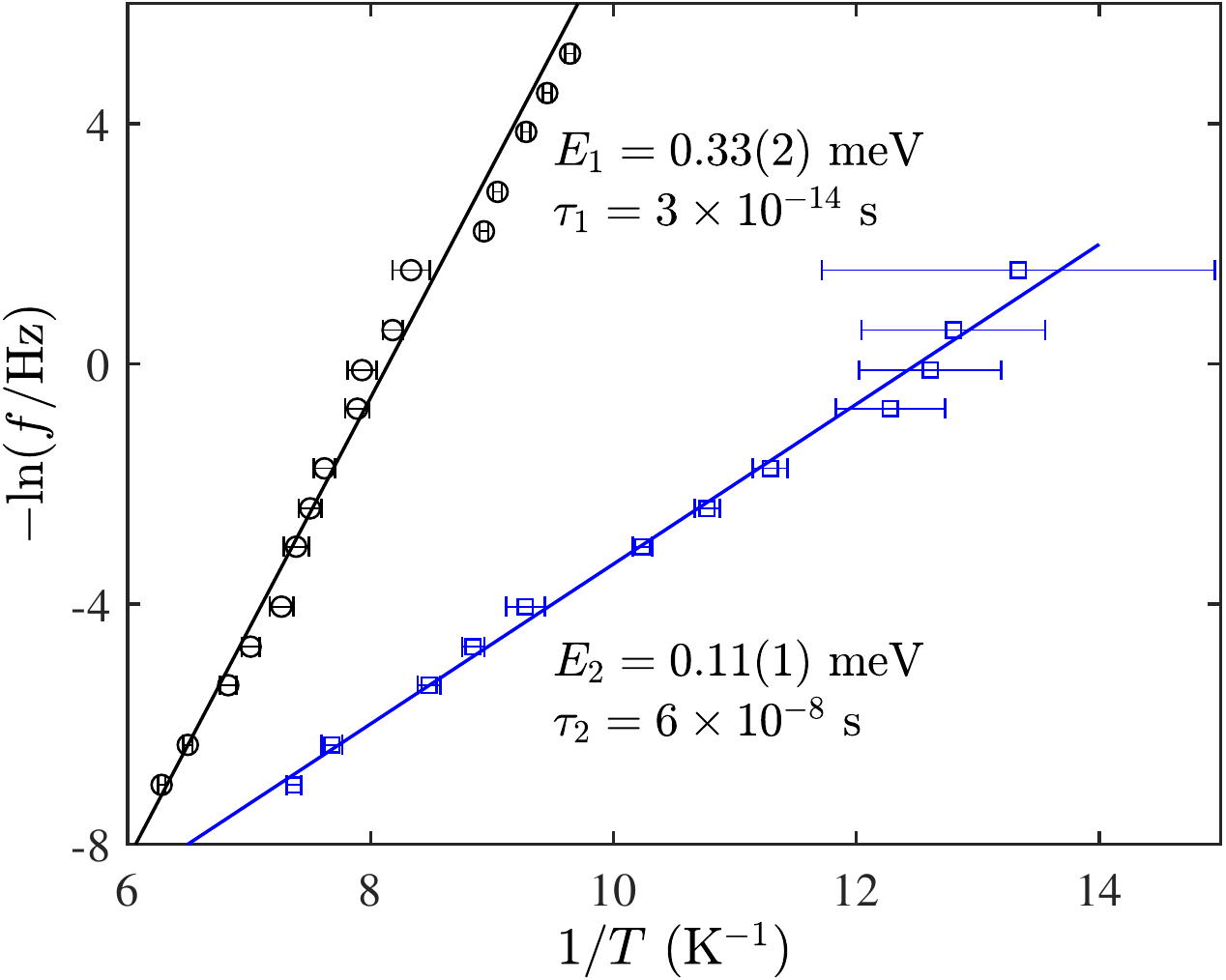}
\caption{Arrhenius plot for the two peaks in $\chi''$ for the single crystal of GGG. The straight lines are fits to the Arrhenius law as described in the text. 
  }
  \label{fig:Fig10}
\end{figure}

The frequency dependence of the peak position of these curves follows an Arrhenius temperature dependence, see Fig.  \ref{fig:Fig10}. 
\begin{equation}
 \tau_{i}=\tau_{0,i}e^{E_{a,i}/k_{B}T},
\label{ArheniusEqn}
\end{equation}
where $i=1,2$, $E_i$ is an activation energy and $\tau_{0,i}=2\pi/f$ the relaxation lifetime of spin vector reversal. The excellent fit to the Arrhenius law indicates that the motion of the spins on this energy scale is governed by  thermal activation across an energy barrier. Two distinct peaks indicate two distinct relaxation processes. 

The fits to the data using the Arrhenius law for each of the two peaks provide activation energies of $E_{a,1}$ = 0.33(2) meV (3.8(3) K) and $E_{a,2}$ = 0.11(1) meV (1.3(2) K) and relaxation lifetimes of $\tau_{0,1}$ = $3\times 10^{-14}$ s and $\tau_{0,2}$ = $6\times 10^{-8}$ s. It should be noted that the uncertainty on $\tau$ is about an order of magnitude. The energy barriers are the same order of magnitude as the excitation energies probed with neutron scattering.

\begin{figure}
\includegraphics[width=0.48\textwidth]{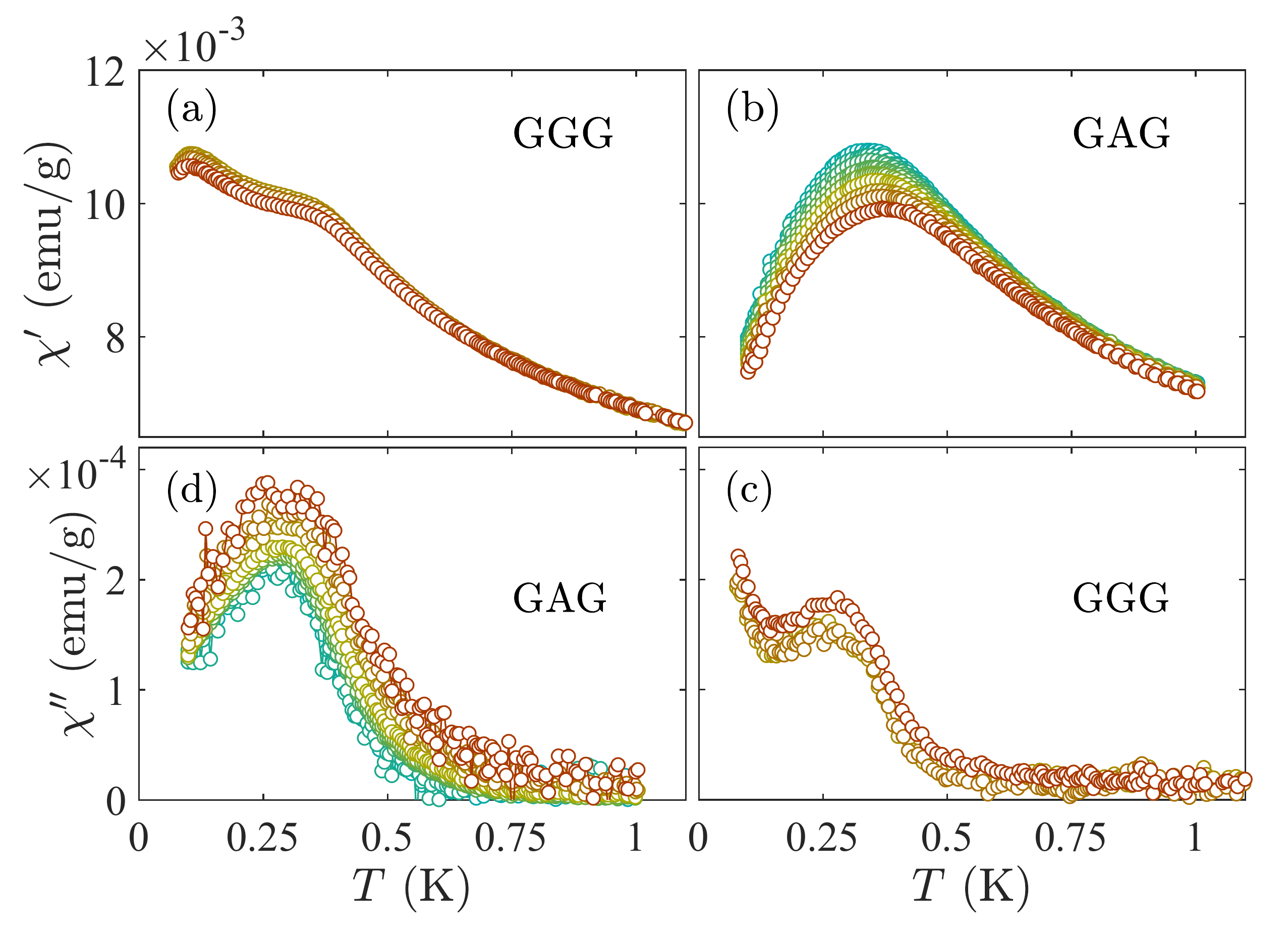}
 \caption{
 Temperature and frequency dependence of the real ($\chi'$) and imaginary ($\chi''$) AC susceptibliity on powders of GGG. The frequencies, $0.57 < f < 211$ Hz, are indicated by the same colors as in Fig.~\ref{fig:Fig9}.
(a) $\chi'$ for powdered GGG.
 (b) $\chi'$ for powdered  GAG.
 (c) $\chi''$ for powdered  GGG.
 (d) $\chi''$ for powdered  GAG. } 
 \label{fig:Fig11}
\end{figure}

The temperature dependence of $\chi'$ for powdered GGG is given in  Fig.~\ref{fig:Fig11}(a) for $f=57$, 111 and 211 Hz. Two anomalies are seen, near 350 mK and 100 mK, as also found previously \cite{Schiffer1995}. The signal is much broader than for the single crystal, and the two anomalies are at different temperatures than the anomaly in $\chi'$ for the single crystal of GGG.

Fig.~\ref{fig:Fig9}(b) shows the temperature dependence of $\chi'$ for powdered GAG for a larger range of frequencies. Here, a single, broad peak is seen around 350 mK. The broadness of the peak indicates a broad distribution of time scales in the system.

The temperature dependence of $\chi''$ is shown in Fig.~\ref{fig:Fig9}(c) for GGG and Fig.~\ref{fig:Fig9}(d) for GAG. In both cases a broad anomaly is seen around 250 mK. The peaks in the powdered data are substantially broader than the single crystal data on GGG, which indicates the existence of a broader distribution of relaxation times in the system. In both compounds the peak temperature of $\chi''$ depends very weakly on frequency. However, fitting to an Arrhenius law,  gives unphysical characteristic times, suggesting that the freezing mechanism is more complex, and cannot be explained with a single energy barrier. Modifications to the Arrhenius law, such as the Vogel-Fulcher law \cite{Shtrikman1981}, can be made to fit the data with sensible parameters, but the correlations between fitted parameters become so large that interpretation of their values is meaningless.
Instead, the frequency dependence of the $\chi'$ peak can be analyzed through the Mydosh parameter $K$, frequently used to describe spin glasses,
\begin{align}
    K=\frac{\Delta T_f}{T_f \Delta \log(f)},
\end{align}
where $T_f$ is the temperature of the $\chi'$ maximum. For the GGG powder, we have insufficient data to determine $K$. For GAG, we obtain $K_\text{GAG}=0.03-0.04$, which falls in the range of insulating spin-glasses, if a bit smaller \cite{Mydosh1993}. Nevertheless, the  broad shape of the $\chi'$ features in GAG and of the ZFC-FC magnetization \cite{Florea2017} seems to preclude a conventional spin-glass transition. In particular, we could not accurately determine  the freezing temperature $T_g$ which would have allowed us to perform dynamical scaling and estimate possible critical exponents. Further measurements, in a larger frequency range are thus needed to conclude on the nature of the freezing in GAG. The broad feature is reminiscent of the heterogeneous freezing scenario proposed for Heisenberg kagom\'e antiferromagnets \cite{Cepas2012}, and thus suggests that multiple time and lengths scales are involved in this relaxation mechanism.  

Our AC susceptibility measurements on GGG and GAG confirm the spin freezing upon cooling below $\sim$175 mK and $\sim$300 mK respectively, with a frequency dependent $\chi'$ and $\chi''$. Most interestingly, the temperature dependence of $\chi''$ for the single crystal of GGG displays an unusual double peaked structure. The frequency dependence of each peak follows an Arrhenius law, thus indicating the presence of two distinct thermally activated processes. Single spin magnetization precession is typically on the ps timescale, e.g. Ref. \onlinecite{Beauvillain1984}, indicating that the first process in GGG, with $\tau_0=3\times10^{-14}$ s, is related to single spin reversal. The second process ($\tau_0=6\times 10^{-8}$ s) is too slow to involve single spins, and is thus probing the dynamics of larger structures, which we can reasonably assume to be the ten-ion loops. These results differ from those by Ghosh et al. \cite{Ghosh2008} who probed the ten-ion looped structure using magnetic hole burning and found temperature independent clusters behaving as quantum objects. In our case the signatures remain thermally driven. 

In contrast to the single crystal measurements on GGG, the temperature dependence of the powdered GAG AC susceptibility displays a very broad single peaked structure, from which we could not extract a characteristic energy scale. Most probably, a wide distribution of correlation lengths is involved in this dynamical process, which is consistent with the smaller strength of the ten-ion loops correlations in GAG compared with GGG.  The larger deviation from the $120^\circ$ spin structure in GAG than in GGG obtained in the diffraction measurements may also explain the differences between the observed slow dynamics.

\section{Summary and conclusion} \label{sec:conclusion}
In summary, we have presented an extensive comparative study of GGG and GAG using neutron diffraction, neutron backscattering and time-of-flight spectroscopy and AC susceptibility. We confirm the existence of long range  director correlations in GAG, and propose that it will be present in all garnets with antiferromagnetic near-neighbor exchange and relatively strong dipolar interaction or local $xy$-anisotropy. The reduced $\mathcal{D}/J_1$ value in GAG as well as possible changes in $J_2$ and $J_3$ result in weaker director correlations.

On the meV scale, GGG and GAG show similar dynamics with three distinct excitations, INS1-3. In particular, the temperature dependence of the excitation energy and lifetime of the INS3 excitation at 0.58 meV is nearly identical in the two compounds, indicating that it is not driven by near neighbor exchange interactions. The other two excitations are reminiscent of spin waves found in the polarized state of GGG.

On the $\mu$eV scale, the spins in GGG fluctuate with a rate that decreases as $T^{1.5}$. The fluctuations are independent of wavevector transfer in the probed range, consistent with the presence of spin diffusion. 
The slowest dynamics probed by AC susceptibility with low frequencies reveal two distinct fluctuation processes in GGG. One of these is associated with single spin fluctuations, while the other is associated to larger structures, most likely the directors.AC susceptibility on powders of GGG and GAG furthermore give insight into the nature of the freezing of the spins that occurs at low temperature. 

We have suggested several directions for further experimental and theoretical work: On the experimental side it would be interesting to ascertain the presence of director correlations in other Gd-based hyperkagome systems, such as  Gd$_3$Te$_2$Li$_3$O$_{12}$. In addition, experimental work to determine the further-neighbor exchange constants in GAG would be valuable. 

Based on our AC susceptibility data we expect the $\mu$eV dynamics of GAG to be different from GGG. This prediction can be straightforwardly tested experimentally on powders of GAG. To determine if the $Q$-independence of the quasi-elastic broadening is caused by spin diffusion or uncorrelated fluctuations, we propose two experiments: First, similar backscattering experiments as presented here, but reaching lower $Q$-values, would reach the regime where the width of the signal decreases linearly with decreasing $Q$ in a spin diffusion model. Such an experiment will be difficult due to the small magnetic signal at the required low $Q$. Second, measurements on an isotope enriched single crystal would reveal any $Q$-dependence on the signal which is currently lost in powder averaging. Unfortunately, $^{160}$Gd is extremely expensive.

With an isotope enriched single crystal the meV dynamics could also be studied further. We expect INS2 to consist of multiple broad bands of excitations that are similar in nature to the excitations found in the field-polarized state. 

Finally, it is also clear that more theoretical work is required. The origin of the INS3 excitation remains unclear, as does the exact nature of the excitations associated with the director correlations. Conventional spin wave theory is not applicable to these systems, but another avenue to approach this problem could be numerical Langevin simulations. 

\begin{acknowledgments}
This work was supported by the Danish Council for Independent Research through DANSCATT. This work is based on neutron scattering experiments performed at Institut Laue-Langevin, France, and at ISIS, UK. 

Experiments  at  the  ISIS  Neutron  and  MuonSource  were  supported  by  beam-time  allocations  from the Science and Technology Facilities Council. 

The work was supported by MAX4ESSFUN as part of the Interreg project ESS\&MAX IV Cross Border Science and Society, Ref. no. KU-001.
HJ acknowledges  funding  from  the  EU  Horizon  2020  programme under the Marie Sklodowska-Curie grant agreement No701647.

We thank Rasmus Tang and Emil Martiny for useful discussions.
\end{acknowledgments}

\clearpage
 

\begin{thebibliography}{48}%
\makeatletter
\providecommand \@ifxundefined [1]{%
 \@ifx{#1\undefined}
}%
\providecommand \@ifnum [1]{%
 \ifnum #1\expandafter \@firstoftwo
 \else \expandafter \@secondoftwo
 \fi
}%
\providecommand \@ifx [1]{%
 \ifx #1\expandafter \@firstoftwo
 \else \expandafter \@secondoftwo
 \fi
}%
\providecommand \natexlab [1]{#1}%
\providecommand \enquote  [1]{``#1''}%
\providecommand \bibnamefont  [1]{#1}%
\providecommand \bibfnamefont [1]{#1}%
\providecommand \citenamefont [1]{#1}%
\providecommand \href@noop [0]{\@secondoftwo}%
\providecommand \href [0]{\begingroup \@sanitize@url \@href}%
\providecommand \@href[1]{\@@startlink{#1}\@@href}%
\providecommand \@@href[1]{\endgroup#1\@@endlink}%
\providecommand \@sanitize@url [0]{\catcode `\\12\catcode `\$12\catcode
  `\&12\catcode `\#12\catcode `\^12\catcode `\_12\catcode `\%12\relax}%
\providecommand \@@startlink[1]{}%
\providecommand \@@endlink[0]{}%
\providecommand \url  [0]{\begingroup\@sanitize@url \@url }%
\providecommand \@url [1]{\endgroup\@href {#1}{\urlprefix }}%
\providecommand \urlprefix  [0]{URL }%
\providecommand \Eprint [0]{\href }%
\providecommand \doibase [0]{http://dx.doi.org/}%
\providecommand \selectlanguage [0]{\@gobble}%
\providecommand \bibinfo  [0]{\@secondoftwo}%
\providecommand \bibfield  [0]{\@secondoftwo}%
\providecommand \translation [1]{[#1]}%
\providecommand \BibitemOpen [0]{}%
\providecommand \bibitemStop [0]{}%
\providecommand \bibitemNoStop [0]{.\EOS\space}%
\providecommand \EOS [0]{\spacefactor3000\relax}%
\providecommand \BibitemShut  [1]{\csname bibitem#1\endcsname}%
\let\auto@bib@innerbib\@empty
\bibitem [{\citenamefont {Fennell}\ \emph {et~al.}(2009)\citenamefont
  {Fennell}, \citenamefont {Deen}, \citenamefont {Wildes}, \citenamefont
  {Schmalzl}, \citenamefont {Prabhakaran}, \citenamefont {Boothroyd},
  \citenamefont {Aldus}, \citenamefont {McMorrow},\ and\ \citenamefont
  {Bramwell}}]{Fennell2009a}%
  \BibitemOpen
  \bibfield  {author} {\bibinfo {author} {\bibfnamefont {T.}~\bibnamefont
  {Fennell}}, \bibinfo {author} {\bibfnamefont {P.~P.}\ \bibnamefont {Deen}},
  \bibinfo {author} {\bibfnamefont {A.~R.}\ \bibnamefont {Wildes}}, \bibinfo
  {author} {\bibfnamefont {K.}~\bibnamefont {Schmalzl}}, \bibinfo {author}
  {\bibfnamefont {D.}~\bibnamefont {Prabhakaran}}, \bibinfo {author}
  {\bibfnamefont {A.~T.}\ \bibnamefont {Boothroyd}}, \bibinfo {author}
  {\bibfnamefont {R.~J.}\ \bibnamefont {Aldus}}, \bibinfo {author}
  {\bibfnamefont {D.~F.}\ \bibnamefont {McMorrow}}, \ and\ \bibinfo {author}
  {\bibfnamefont {S.~T.}\ \bibnamefont {Bramwell}},\ }\bibfield  {title}
  {\enquote {\bibinfo {title} {{Magnetic Coulomb phase in the spin ice
  Ho$_2$Ti$_2$O$_7$.}}}\ }\href {\doibase 10.1126/science.1177582} {\bibfield
  {journal} {\bibinfo  {journal} {Science}\ }\textbf {\bibinfo {volume}
  {326}},\ \bibinfo {pages} {415} (\bibinfo {year} {2009})}\BibitemShut
  {NoStop}%
\bibitem [{\citenamefont {Morris}\ \emph {et~al.}(2009)\citenamefont {Morris},
  \citenamefont {Tennant}, \citenamefont {Grigera}, \citenamefont {Klemke},
  \citenamefont {Castelnovo}, \citenamefont {Moessner}, \citenamefont
  {Czternasty}, \citenamefont {Meissner}, \citenamefont {Rule}, \citenamefont
  {Hoffmann}, \citenamefont {Kiefer}, \citenamefont {Gerischer}, \citenamefont
  {Slobinsky},\ and\ \citenamefont {Perry}}]{Morris2009}%
  \BibitemOpen
  \bibfield  {author} {\bibinfo {author} {\bibfnamefont {D.~J.~P.}\
  \bibnamefont {Morris}}, \bibinfo {author} {\bibfnamefont {D.~A.}\
  \bibnamefont {Tennant}}, \bibinfo {author} {\bibfnamefont {S.~A.}\
  \bibnamefont {Grigera}}, \bibinfo {author} {\bibfnamefont {B.}~\bibnamefont
  {Klemke}}, \bibinfo {author} {\bibfnamefont {C.}~\bibnamefont {Castelnovo}},
  \bibinfo {author} {\bibfnamefont {R.}~\bibnamefont {Moessner}}, \bibinfo
  {author} {\bibfnamefont {C.}~\bibnamefont {Czternasty}}, \bibinfo {author}
  {\bibfnamefont {M.}~\bibnamefont {Meissner}}, \bibinfo {author}
  {\bibfnamefont {K.~C.}\ \bibnamefont {Rule}}, \bibinfo {author}
  {\bibfnamefont {J.}~\bibnamefont {Hoffmann}}, \bibinfo {author}
  {\bibfnamefont {K.}~\bibnamefont {Kiefer}}, \bibinfo {author} {\bibfnamefont
  {S.}~\bibnamefont {Gerischer}}, \bibinfo {author} {\bibfnamefont
  {D.}~\bibnamefont {Slobinsky}}, \ and\ \bibinfo {author} {\bibfnamefont
  {R.~S.}\ \bibnamefont {Perry}},\ }\bibfield  {title} {\enquote {\bibinfo
  {title} {{Dirac Strings and Magnetic Monopoles in the Spin Ice
  Dy$_2$Ti$_2$O$_7$}},}\ }\href {\doibase 10.1126/science.1178868} {\bibfield
  {journal} {\bibinfo  {journal} {Science}\ }\textbf {\bibinfo {volume}
  {326}},\ \bibinfo {pages} {411} (\bibinfo {year} {2009})}\BibitemShut
  {NoStop}%
\bibitem [{\citenamefont {Castelnovo}\ \emph {et~al.}(2007)\citenamefont
  {Castelnovo}, \citenamefont {Moessner},\ and\ \citenamefont
  {Sondhi}}]{Castelnovo2008}%
  \BibitemOpen
  \bibfield  {author} {\bibinfo {author} {\bibfnamefont {C.}~\bibnamefont
  {Castelnovo}}, \bibinfo {author} {\bibfnamefont {R.}~\bibnamefont
  {Moessner}}, \ and\ \bibinfo {author} {\bibfnamefont {S.~L.}\ \bibnamefont
  {Sondhi}},\ }\bibfield  {title} {\enquote {\bibinfo {title} {{Magnetic
  monopoles in spin ice}},}\ }\href {\doibase 10.1038/nature06433} {\bibfield
  {journal} {\bibinfo  {journal} {Nature}\ }\textbf {\bibinfo {volume} {451}},\
  \bibinfo {pages} {42} (\bibinfo {year} {2007})}\BibitemShut {NoStop}%
\bibitem [{\citenamefont {Giblin}\ \emph {et~al.}(2011)\citenamefont {Giblin},
  \citenamefont {Bramwell}, \citenamefont {Holdsworth}, \citenamefont
  {Prabhakaran},\ and\ \citenamefont {Terry}}]{Giblin2011}%
  \BibitemOpen
  \bibfield  {author} {\bibinfo {author} {\bibfnamefont {S.~R.}\ \bibnamefont
  {Giblin}}, \bibinfo {author} {\bibfnamefont {S.~T.}\ \bibnamefont
  {Bramwell}}, \bibinfo {author} {\bibfnamefont {P.~C.~W.}\ \bibnamefont
  {Holdsworth}}, \bibinfo {author} {\bibfnamefont {D.}~\bibnamefont
  {Prabhakaran}}, \ and\ \bibinfo {author} {\bibfnamefont {I.}~\bibnamefont
  {Terry}},\ }\bibfield  {title} {\enquote {\bibinfo {title} {{Creation and
  measurement of long-lived magnetic monopole currents in spin ice}},}\ }\href
  {\doibase 10.1038/nphys1896} {\bibfield  {journal} {\bibinfo  {journal}
  {Nature Physics}\ }\textbf {\bibinfo {volume} {7}},\ \bibinfo {pages} {252}
  (\bibinfo {year} {2011})}\BibitemShut {NoStop}%
\bibitem [{\citenamefont {Dusad}\ \emph {et~al.}(2019)\citenamefont {Dusad},
  \citenamefont {Kirschner}, \citenamefont {Hoke}, \citenamefont {Roberts},
  \citenamefont {Eyal}, \citenamefont {Flicker}, \citenamefont {Luke},
  \citenamefont {Blundell},\ and\ \citenamefont {Davis}}]{Dusad2019}%
  \BibitemOpen
  \bibfield  {author} {\bibinfo {author} {\bibfnamefont {R.}~\bibnamefont
  {Dusad}}, \bibinfo {author} {\bibfnamefont {F.~K.~K.}\ \bibnamefont
  {Kirschner}}, \bibinfo {author} {\bibfnamefont {J.~C.}\ \bibnamefont {Hoke}},
  \bibinfo {author} {\bibfnamefont {B.~R.}\ \bibnamefont {Roberts}}, \bibinfo
  {author} {\bibfnamefont {A.}~\bibnamefont {Eyal}}, \bibinfo {author}
  {\bibfnamefont {F.}~\bibnamefont {Flicker}}, \bibinfo {author} {\bibfnamefont
  {G.~M.}\ \bibnamefont {Luke}}, \bibinfo {author} {\bibfnamefont {S.~J.}\
  \bibnamefont {Blundell}}, \ and\ \bibinfo {author} {\bibfnamefont {J.~C.~S.}\
  \bibnamefont {Davis}},\ }\bibfield  {title} {\enquote {\bibinfo {title}
  {{Magnetic monopole noise}},}\ }\href {\doibase 10.1038/s41586-019-1358-1}
  {\bibfield  {journal} {\bibinfo  {journal} {Nature}\ }\textbf {\bibinfo
  {volume} {571}},\ \bibinfo {pages} {234} (\bibinfo {year}
  {2019})}\BibitemShut {NoStop}%
\bibitem [{\citenamefont {Paddison}\ \emph {et~al.}(2015)\citenamefont
  {Paddison}, \citenamefont {Jacobsen}, \citenamefont {Petrenko}, \citenamefont
  {Fernandez-Diaz}, \citenamefont {Deen},\ and\ \citenamefont
  {Goodwin}}]{Paddison2015a}%
  \BibitemOpen
  \bibfield  {author} {\bibinfo {author} {\bibfnamefont {J.~A.~M.}\
  \bibnamefont {Paddison}}, \bibinfo {author} {\bibfnamefont {H.}~\bibnamefont
  {Jacobsen}}, \bibinfo {author} {\bibfnamefont {O.~A.}\ \bibnamefont
  {Petrenko}}, \bibinfo {author} {\bibfnamefont {M.~T.}\ \bibnamefont
  {Fernandez-Diaz}}, \bibinfo {author} {\bibfnamefont {P.~P.}\ \bibnamefont
  {Deen}}, \ and\ \bibinfo {author} {\bibfnamefont {A.~L.}\ \bibnamefont
  {Goodwin}},\ }\bibfield  {title} {\enquote {\bibinfo {title} {{Hidden order
  in spin-liquid Gd$_3$Ga$_5$O$_{12}$}},}\ }\href {\doibase
  10.1126/science.aaa5326} {\bibfield  {journal} {\bibinfo  {journal}
  {Science}\ }\textbf {\bibinfo {volume} {350}},\ \bibinfo {pages} {179}
  (\bibinfo {year} {2015})}\BibitemShut {NoStop}%
\bibitem [{\citenamefont {Lee}\ \emph {et~al.}(2002)\citenamefont {Lee},
  \citenamefont {Broholm}, \citenamefont {Ratcliff}, \citenamefont
  {Gasparovic}, \citenamefont {Huang}, \citenamefont {Kim},\ and\ \citenamefont
  {Cheong}}]{Lee2002}%
  \BibitemOpen
  \bibfield  {author} {\bibinfo {author} {\bibfnamefont {S.}~\bibnamefont
  {Lee}}, \bibinfo {author} {\bibfnamefont {C.}~\bibnamefont {Broholm}},
  \bibinfo {author} {\bibfnamefont {W.}~\bibnamefont {Ratcliff}}, \bibinfo
  {author} {\bibfnamefont {G.}~\bibnamefont {Gasparovic}}, \bibinfo {author}
  {\bibfnamefont {Q.}~\bibnamefont {Huang}}, \bibinfo {author} {\bibfnamefont
  {T.~H.}\ \bibnamefont {Kim}}, \ and\ \bibinfo {author} {\bibfnamefont
  {S.}~\bibnamefont {Cheong}},\ }\bibfield  {title} {\enquote {\bibinfo {title}
  {{Emergent excitations in a geometrically frustrated magnet.}}}\ }\href
  {\doibase 10.1038/nature00964} {\bibfield  {journal} {\bibinfo  {journal}
  {Nature}\ }\textbf {\bibinfo {volume} {418}},\ \bibinfo {pages} {856}
  (\bibinfo {year} {2002})}\BibitemShut {NoStop}%
\bibitem [{\citenamefont {Ghosh}\ \emph {et~al.}(2008)\citenamefont {Ghosh},
  \citenamefont {Rosenbaum},\ and\ \citenamefont {Aeppli}}]{Ghosh2008}%
  \BibitemOpen
  \bibfield  {author} {\bibinfo {author} {\bibfnamefont {S.}~\bibnamefont
  {Ghosh}}, \bibinfo {author} {\bibfnamefont {T.~F.}\ \bibnamefont
  {Rosenbaum}}, \ and\ \bibinfo {author} {\bibfnamefont {G.}~\bibnamefont
  {Aeppli}},\ }\bibfield  {title} {\enquote {\bibinfo {title} {{Macroscopic
  signature of protected spins in a dense frustrated magnet}},}\ }\href
  {\doibase 10.1103/PhysRevLett.101.157205} {\bibfield  {journal} {\bibinfo
  {journal} {Phys. Rev. Lett.}\ }\textbf {\bibinfo {volume} {101}},\ \bibinfo
  {pages} {157205} (\bibinfo {year} {2008})}\BibitemShut {NoStop}%
\bibitem [{\citenamefont {Deen}\ \emph {et~al.}(2010)\citenamefont {Deen},
  \citenamefont {Petrenko}, \citenamefont {Balakrishnan}, \citenamefont
  {Rainford}, \citenamefont {Ritter}, \citenamefont {Capogna}, \citenamefont
  {Mutka},\ and\ \citenamefont {Fennell}}]{Deen2010}%
  \BibitemOpen
  \bibfield  {author} {\bibinfo {author} {\bibfnamefont {P.~P.}\ \bibnamefont
  {Deen}}, \bibinfo {author} {\bibfnamefont {O.~A.}\ \bibnamefont {Petrenko}},
  \bibinfo {author} {\bibfnamefont {G.}~\bibnamefont {Balakrishnan}}, \bibinfo
  {author} {\bibfnamefont {B.~D.}\ \bibnamefont {Rainford}}, \bibinfo {author}
  {\bibfnamefont {C.}~\bibnamefont {Ritter}}, \bibinfo {author} {\bibfnamefont
  {L.}~\bibnamefont {Capogna}}, \bibinfo {author} {\bibfnamefont
  {H.}~\bibnamefont {Mutka}}, \ and\ \bibinfo {author} {\bibfnamefont
  {T.}~\bibnamefont {Fennell}},\ }\bibfield  {title} {\enquote {\bibinfo
  {title} {{Spin dynamics in the hyperkagome compound Gd$_3$Ga$_5$O$_{12}$}},}\
  }\href {\doibase 10.1103/PhysRevB.82.174408} {\bibfield  {journal} {\bibinfo
  {journal} {Phys. Rev. B}\ }\textbf {\bibinfo {volume} {82}},\ \bibinfo
  {pages} {174408} (\bibinfo {year} {2010})}\BibitemShut {NoStop}%
\bibitem [{\citenamefont {Dunsiger}\ \emph {et~al.}(2000)\citenamefont
  {Dunsiger}, \citenamefont {Gardner}, \citenamefont {Chakhalian},
  \citenamefont {Cornelius}, \citenamefont {Jaime}, \citenamefont {Kiefl},
  \citenamefont {Movshovich}, \citenamefont {MacFarlane}, \citenamefont
  {Miller}, \citenamefont {Sonier},\ and\ \citenamefont
  {Gaulin}}]{Dunsiger2000}%
  \BibitemOpen
  \bibfield  {author} {\bibinfo {author} {\bibfnamefont {S.~R.}\ \bibnamefont
  {Dunsiger}}, \bibinfo {author} {\bibfnamefont {J.~S.}\ \bibnamefont
  {Gardner}}, \bibinfo {author} {\bibfnamefont {J.~A.}\ \bibnamefont
  {Chakhalian}}, \bibinfo {author} {\bibfnamefont {A.~L.}\ \bibnamefont
  {Cornelius}}, \bibinfo {author} {\bibfnamefont {M.}~\bibnamefont {Jaime}},
  \bibinfo {author} {\bibfnamefont {R.~F.}\ \bibnamefont {Kiefl}}, \bibinfo
  {author} {\bibfnamefont {R.}~\bibnamefont {Movshovich}}, \bibinfo {author}
  {\bibfnamefont {W.~A.}\ \bibnamefont {MacFarlane}}, \bibinfo {author}
  {\bibfnamefont {R.~I.}\ \bibnamefont {Miller}}, \bibinfo {author}
  {\bibfnamefont {J.~E.}\ \bibnamefont {Sonier}}, \ and\ \bibinfo {author}
  {\bibfnamefont {B.~D.}\ \bibnamefont {Gaulin}},\ }\bibfield  {title}
  {\enquote {\bibinfo {title} {{Low temperature spin dynamics of the
  geometrically frustrated antiferromagnetic garnet Gd$_3$Ga$_5$O$_{12}$}},}\
  }\href {\doibase 10.1103/PhysRevLett.85.3504} {\bibfield  {journal} {\bibinfo
   {journal} {Phys. Rev. Lett.}\ }\textbf {\bibinfo {volume} {85}},\ \bibinfo
  {pages} {3504} (\bibinfo {year} {2000})}\BibitemShut {NoStop}%
\bibitem [{\citenamefont {Marshall}\ \emph {et~al.}(2002)\citenamefont
  {Marshall}, \citenamefont {Blundell}, \citenamefont {Pratt}, \citenamefont
  {Husmann}, \citenamefont {Steer}, \citenamefont {Coldea}, \citenamefont
  {Hayes},\ and\ \citenamefont {Ward}}]{Marshall2002}%
  \BibitemOpen
  \bibfield  {author} {\bibinfo {author} {\bibfnamefont {I.~M.}\ \bibnamefont
  {Marshall}}, \bibinfo {author} {\bibfnamefont {S.~J.}\ \bibnamefont
  {Blundell}}, \bibinfo {author} {\bibfnamefont {F.~L.}\ \bibnamefont {Pratt}},
  \bibinfo {author} {\bibfnamefont {A.}~\bibnamefont {Husmann}}, \bibinfo
  {author} {\bibfnamefont {C.~A.}\ \bibnamefont {Steer}}, \bibinfo {author}
  {\bibfnamefont {A.~I.}\ \bibnamefont {Coldea}}, \bibinfo {author}
  {\bibfnamefont {W.}~\bibnamefont {Hayes}}, \ and\ \bibinfo {author}
  {\bibfnamefont {R.~C.~C.}\ \bibnamefont {Ward}},\ }\bibfield  {title}
  {\enquote {\bibinfo {title} {{A muon-spin relaxation ($\mu$SR) study of the
  geometrically frustrated magnets Gd$_3$Ga$_5$O$_{12}$ and ZnCr$_2$O$_4$}},}\
  }\href {\doibase 10.1016/j.jlumin.2014.11.010} {\bibfield  {journal}
  {\bibinfo  {journal} {J. Phys. Condens. Matter}\ }\textbf {\bibinfo {volume}
  {14}},\ \bibinfo {pages} {L157} (\bibinfo {year} {2002})}\BibitemShut
  {NoStop}%
\bibitem [{\citenamefont {Bonville}\ \emph {et~al.}(2004)\citenamefont
  {Bonville}, \citenamefont {Hodges}, \citenamefont {Sanchez},\ and\
  \citenamefont {Vulliet}}]{Bonville2004}%
  \BibitemOpen
  \bibfield  {author} {\bibinfo {author} {\bibfnamefont {P.}~\bibnamefont
  {Bonville}}, \bibinfo {author} {\bibfnamefont {J.~A.}\ \bibnamefont
  {Hodges}}, \bibinfo {author} {\bibfnamefont {J.~P.}\ \bibnamefont {Sanchez}},
  \ and\ \bibinfo {author} {\bibfnamefont {P.}~\bibnamefont {Vulliet}},\
  }\bibfield  {title} {\enquote {\bibinfo {title} {{Planar spin fluctuations
  with a quadratic thermal dependence rate in spin liquid
  Gd$_3$Ga$_5$O$_{12}$}},}\ }\href {\doibase 10.1103/PhysRevLett.92.167202}
  {\bibfield  {journal} {\bibinfo  {journal} {Phys. Rev. Lett.}\ }\textbf
  {\bibinfo {volume} {92}},\ \bibinfo {pages} {167202} (\bibinfo {year}
  {2004})}\BibitemShut {NoStop}%
\bibitem [{\citenamefont {d'Ambrumenil}\ \emph {et~al.}(2015)\citenamefont
  {d'Ambrumenil}, \citenamefont {Petrenko}, \citenamefont {Mutka},\ and\
  \citenamefont {Deen}}]{DAmbrumenil2015}%
  \BibitemOpen
  \bibfield  {author} {\bibinfo {author} {\bibfnamefont {N.}~\bibnamefont
  {d'Ambrumenil}}, \bibinfo {author} {\bibfnamefont {O.~A.}\ \bibnamefont
  {Petrenko}}, \bibinfo {author} {\bibfnamefont {H.}~\bibnamefont {Mutka}}, \
  and\ \bibinfo {author} {\bibfnamefont {P.~P.}\ \bibnamefont {Deen}},\
  }\bibfield  {title} {\enquote {\bibinfo {title} {{Dispersionless Spin Waves
  and Underlying Field-Induced Magnetic Order in Gadolinium Gallium Garnet}},}\
  }\href {\doibase 10.1103/PhysRevLett.114.227203} {\bibfield  {journal}
  {\bibinfo  {journal} {Phys. Rev. Lett.}\ }\textbf {\bibinfo {volume} {114}},\
  \bibinfo {pages} {227203} (\bibinfo {year} {2015})}\BibitemShut {NoStop}%
\bibitem [{\citenamefont {Lefran{\c{c}}ois}\ \emph {et~al.}(2019)\citenamefont
  {Lefran{\c{c}}ois}, \citenamefont {Mangin-Thro}, \citenamefont {Lhotel},
  \citenamefont {Robert}, \citenamefont {Petit}, \citenamefont {Cathelin},
  \citenamefont {Fischer}, \citenamefont {Colin}, \citenamefont {Damay},
  \citenamefont {Ollivier}, \citenamefont {Lejay}, \citenamefont {Chapon},
  \citenamefont {Simonet},\ and\ \citenamefont {Ballou}}]{Lefrancois2019}%
  \BibitemOpen
  \bibfield  {author} {\bibinfo {author} {\bibfnamefont {E.}~\bibnamefont
  {Lefran{\c{c}}ois}}, \bibinfo {author} {\bibfnamefont {L.}~\bibnamefont
  {Mangin-Thro}}, \bibinfo {author} {\bibfnamefont {E.}~\bibnamefont {Lhotel}},
  \bibinfo {author} {\bibfnamefont {J.}~\bibnamefont {Robert}}, \bibinfo
  {author} {\bibfnamefont {S.}~\bibnamefont {Petit}}, \bibinfo {author}
  {\bibfnamefont {V.}~\bibnamefont {Cathelin}}, \bibinfo {author}
  {\bibfnamefont {H.~E.}\ \bibnamefont {Fischer}}, \bibinfo {author}
  {\bibfnamefont {C.~V.}\ \bibnamefont {Colin}}, \bibinfo {author}
  {\bibfnamefont {F.}~\bibnamefont {Damay}}, \bibinfo {author} {\bibfnamefont
  {J.}~\bibnamefont {Ollivier}}, \bibinfo {author} {\bibfnamefont
  {P.}~\bibnamefont {Lejay}}, \bibinfo {author} {\bibfnamefont {L.~C.}\
  \bibnamefont {Chapon}}, \bibinfo {author} {\bibfnamefont {V.}~\bibnamefont
  {Simonet}}, \ and\ \bibinfo {author} {\bibfnamefont {R.}~\bibnamefont
  {Ballou}},\ }\bibfield  {title} {\enquote {\bibinfo {title} {{Spin decoupling
  under a staggered field in the Gd$_2$Ir$_2$O$_7$ pyrochlore}},}\ }\href
  {\doibase 10.1103/PhysRevB.99.060401} {\bibfield  {journal} {\bibinfo
  {journal} {Phys. Rev. B}\ }\textbf {\bibinfo {volume} {99}},\ \bibinfo
  {pages} {060401(R)} (\bibinfo {year} {2019})}\BibitemShut {NoStop}%
\bibitem [{\citenamefont {Kinney}\ and\ \citenamefont
  {Wolf}(1979)}]{Kinney1979}%
  \BibitemOpen
  \bibfield  {author} {\bibinfo {author} {\bibfnamefont {W.~I.}\ \bibnamefont
  {Kinney}}\ and\ \bibinfo {author} {\bibfnamefont {W.~P.}\ \bibnamefont
  {Wolf}},\ }\bibfield  {title} {\enquote {\bibinfo {title} {{Magnetic
  interactions and short-range order in gadolinium gallium garnet}},}\ }\href
  {\doibase 10.1063/1.326954} {\bibfield  {journal} {\bibinfo  {journal}
  {Journal of Applied Physics}\ }\textbf {\bibinfo {volume} {50}},\ \bibinfo
  {pages} {2115} (\bibinfo {year} {1979})}\BibitemShut {NoStop}%
\bibitem [{\citenamefont {Petrenko}\ \emph {et~al.}(1998)\citenamefont
  {Petrenko}, \citenamefont {Ritter}, \citenamefont {Yethiraj},\ and\
  \citenamefont {{McK Paul}}}]{Petrenko1998}%
  \BibitemOpen
  \bibfield  {author} {\bibinfo {author} {\bibfnamefont {O.~A.}\ \bibnamefont
  {Petrenko}}, \bibinfo {author} {\bibfnamefont {C.}~\bibnamefont {Ritter}},
  \bibinfo {author} {\bibfnamefont {M.}~\bibnamefont {Yethiraj}}, \ and\
  \bibinfo {author} {\bibfnamefont {D.}~\bibnamefont {{McK Paul}}},\ }\bibfield
   {title} {\enquote {\bibinfo {title} {{Investigation of the Low-Temperature
  Spin-Liquid Behavior of the Frustrated Magnet Gadolinium Gallium Garnet}},}\
  }\href {\doibase 10.1103/PhysRevLett.80.4570} {\bibfield  {journal} {\bibinfo
   {journal} {Phys. Rev. Lett.}\ }\textbf {\bibinfo {volume} {80}},\ \bibinfo
  {pages} {4570} (\bibinfo {year} {1998})}\BibitemShut {NoStop}%
\bibitem [{\citenamefont {Schiffer}\ \emph {et~al.}(1994)\citenamefont
  {Schiffer}, \citenamefont {Ramirez}, \citenamefont {Huse},\ and\
  \citenamefont {Valentino}}]{Schiffer1994}%
  \BibitemOpen
  \bibfield  {author} {\bibinfo {author} {\bibfnamefont {P.}~\bibnamefont
  {Schiffer}}, \bibinfo {author} {\bibfnamefont {A.~P.}\ \bibnamefont
  {Ramirez}}, \bibinfo {author} {\bibfnamefont {D.~A.}\ \bibnamefont {Huse}}, \
  and\ \bibinfo {author} {\bibfnamefont {A.~J.}\ \bibnamefont {Valentino}},\
  }\bibfield  {title} {\enquote {\bibinfo {title} {{Investigation of the field
  induced antiferromagnetic phase transition in the frustrated magnet:
  Gadolinium gallium garnet}},}\ }\href {\doibase 10.1103/PhysRevLett.73.2500}
  {\bibfield  {journal} {\bibinfo  {journal} {Phys. Rev. Lett.}\ }\textbf
  {\bibinfo {volume} {73}},\ \bibinfo {pages} {2500} (\bibinfo {year}
  {1994})}\BibitemShut {NoStop}%
\bibitem [{\citenamefont {Quilliam}\ \emph {et~al.}(2013)\citenamefont
  {Quilliam}, \citenamefont {Meng}, \citenamefont {Craig}, \citenamefont
  {Corruccini}, \citenamefont {Balakrishnan}, \citenamefont {Petrenko},
  \citenamefont {Gomez}, \citenamefont {Kycia}, \citenamefont {Gingras},\ and\
  \citenamefont {Kycia}}]{Quilliam2013}%
  \BibitemOpen
  \bibfield  {author} {\bibinfo {author} {\bibfnamefont {J.~A.}\ \bibnamefont
  {Quilliam}}, \bibinfo {author} {\bibfnamefont {S.}~\bibnamefont {Meng}},
  \bibinfo {author} {\bibfnamefont {H.~A.}\ \bibnamefont {Craig}}, \bibinfo
  {author} {\bibfnamefont {L.~R.}\ \bibnamefont {Corruccini}}, \bibinfo
  {author} {\bibfnamefont {G.}~\bibnamefont {Balakrishnan}}, \bibinfo {author}
  {\bibfnamefont {O.~A.}\ \bibnamefont {Petrenko}}, \bibinfo {author}
  {\bibfnamefont {A.}~\bibnamefont {Gomez}}, \bibinfo {author} {\bibfnamefont
  {S.~W.}\ \bibnamefont {Kycia}}, \bibinfo {author} {\bibfnamefont {M.~J.~P.}\
  \bibnamefont {Gingras}}, \ and\ \bibinfo {author} {\bibfnamefont {J.~B.}\
  \bibnamefont {Kycia}},\ }\bibfield  {title} {\enquote {\bibinfo {title}
  {{Juxtaposition of spin freezing and long range order in a series of
  geometrically frustrated antiferromagnetic gadolinium garnets}},}\ }\href
  {\doibase 10.1103/PhysRevB.87.174421} {\bibfield  {journal} {\bibinfo
  {journal} {Phys. Rev. B}\ }\textbf {\bibinfo {volume} {87}},\ \bibinfo
  {pages} {174421} (\bibinfo {year} {2013})}\BibitemShut {NoStop}%
\bibitem [{\citenamefont {Florea}\ \emph {et~al.}(2017)\citenamefont {Florea},
  \citenamefont {Lhotel}, \citenamefont {Jacobsen}, \citenamefont {Knee},\ and\
  \citenamefont {Deen}}]{Florea2017}%
  \BibitemOpen
  \bibfield  {author} {\bibinfo {author} {\bibfnamefont {O.}~\bibnamefont
  {Florea}}, \bibinfo {author} {\bibfnamefont {E.}~\bibnamefont {Lhotel}},
  \bibinfo {author} {\bibfnamefont {H.}~\bibnamefont {Jacobsen}}, \bibinfo
  {author} {\bibfnamefont {C.~S.}\ \bibnamefont {Knee}}, \ and\ \bibinfo
  {author} {\bibfnamefont {P.~P.}\ \bibnamefont {Deen}},\ }\bibfield  {title}
  {\enquote {\bibinfo {title} {{Absence of magnetic ordering and field-induced
  phase diagram in the gadolinium aluminum garnet}},}\ }\href {\doibase
  10.1103/PhysRevB.96.220413} {\bibfield  {journal} {\bibinfo  {journal} {Phys.
  Rev. B}\ }\textbf {\bibinfo {volume} {96}},\ \bibinfo {pages} {220413(R)}
  (\bibinfo {year} {2017})}\BibitemShut {NoStop}%
\bibitem [{\citenamefont {Yavors'kii}\ \emph {et~al.}(2006)\citenamefont
  {Yavors'kii}, \citenamefont {Enjalran},\ and\ \citenamefont
  {Gingras}}]{Yavorskii2006}%
  \BibitemOpen
  \bibfield  {author} {\bibinfo {author} {\bibfnamefont {T.}~\bibnamefont
  {Yavors'kii}}, \bibinfo {author} {\bibfnamefont {M.}~\bibnamefont
  {Enjalran}}, \ and\ \bibinfo {author} {\bibfnamefont {M.~J.~P.}\ \bibnamefont
  {Gingras}},\ }\bibfield  {title} {\enquote {\bibinfo {title} {{Spin
  hamiltonian, competing small energy scales, and incommensurate long-range
  order in the highly frustrated Gd$_3$Ga$_5$O$_{12}$ garnet
  antiferromagnet}},}\ }\href {\doibase 10.1103/PhysRevLett.97.267203}
  {\bibfield  {journal} {\bibinfo  {journal} {Phys. Rev. Lett.}\ }\textbf
  {\bibinfo {volume} {97}},\ \bibinfo {pages} {267203} (\bibinfo {year}
  {2006})}\BibitemShut {NoStop}%
\bibitem [{\citenamefont {Deen}\ \emph
  {et~al.}(2015{\natexlab{a}})\citenamefont {Deen}, \citenamefont {Florea},
  \citenamefont {Lhotel},\ and\ \citenamefont {Jacobsen}}]{Deen2015}%
  \BibitemOpen
  \bibfield  {author} {\bibinfo {author} {\bibfnamefont {P.~P.}\ \bibnamefont
  {Deen}}, \bibinfo {author} {\bibfnamefont {O.}~\bibnamefont {Florea}},
  \bibinfo {author} {\bibfnamefont {E.}~\bibnamefont {Lhotel}}, \ and\ \bibinfo
  {author} {\bibfnamefont {H.}~\bibnamefont {Jacobsen}},\ }\bibfield  {title}
  {\enquote {\bibinfo {title} {{Updating the phase diagram of the archetypal
  frustrated magnet Gd$_3$Ga$_5$O$_{12}$}},}\ }\href {\doibase
  10.1103/PhysRevB.91.014419} {\bibfield  {journal} {\bibinfo  {journal} {Phys.
  Rev. B}\ }\textbf {\bibinfo {volume} {91}},\ \bibinfo {pages} {014419}
  (\bibinfo {year} {2015}{\natexlab{a}})}\BibitemShut {NoStop}%
\bibitem [{\citenamefont {Petrenko}\ \emph {et~al.}(1997)\citenamefont
  {Petrenko}, \citenamefont {Ritter}, \citenamefont {Yethiraj},\ and\
  \citenamefont {{McK Paul}}}]{Petrenko1998a}%
  \BibitemOpen
  \bibfield  {author} {\bibinfo {author} {\bibfnamefont {O.~A.}\ \bibnamefont
  {Petrenko}}, \bibinfo {author} {\bibfnamefont {C.}~\bibnamefont {Ritter}},
  \bibinfo {author} {\bibfnamefont {M.}~\bibnamefont {Yethiraj}}, \ and\
  \bibinfo {author} {\bibfnamefont {D.}~\bibnamefont {{McK Paul}}},\ }\bibfield
   {title} {\enquote {\bibinfo {title} {{Spin-liquid behavior of the gadolinium
  gallium garnet}},}\ }\href {\doibase 10.1103/PhysRevLett.80.4570} {\bibfield
  {journal} {\bibinfo  {journal} {Physica B}\ }\textbf {\bibinfo {volume}
  {241-243}},\ \bibinfo {pages} {727} (\bibinfo {year} {1997})}\BibitemShut
  {NoStop}%
\bibitem [{\citenamefont {Deen}\ \emph
  {et~al.}(2015{\natexlab{b}})\citenamefont {Deen}, \citenamefont {Florea},
  \citenamefont {Paul}, \citenamefont {Jacobsen}, \citenamefont {Khaplanov},
  \citenamefont {Knee}, \citenamefont {Lhotel},\ and\ \citenamefont
  {Wildes}}]{D7_data}%
  \BibitemOpen
  \bibfield  {author} {\bibinfo {author} {\bibfnamefont {P.~P.}\ \bibnamefont
  {Deen}}, \bibinfo {author} {\bibfnamefont {O.}~\bibnamefont {Florea}},
  \bibinfo {author} {\bibfnamefont {H.}~\bibnamefont {Paul}}, \bibinfo {author}
  {\bibfnamefont {H.}~\bibnamefont {Jacobsen}}, \bibinfo {author}
  {\bibfnamefont {A.}~\bibnamefont {Khaplanov}}, \bibinfo {author}
  {\bibfnamefont {C.}~\bibnamefont {Knee}}, \bibinfo {author} {\bibfnamefont
  {E.}~\bibnamefont {Lhotel}}, \ and\ \bibinfo {author} {\bibfnamefont
  {A.}~\bibnamefont {Wildes}},\ }\href@noop {} {\enquote {\bibinfo {title}
  {Magnetic frustration in 3d hyperkagome compounds},}\ }\bibinfo
  {howpublished} {Institut Laue-Langevin (ILL), doi:10.5291/ILL-DATA.5-32-805}
  (\bibinfo {year} {2015}{\natexlab{b}})\BibitemShut {NoStop}%
\bibitem [{\citenamefont {Deen}\ \emph
  {et~al.}(2014{\natexlab{a}})\citenamefont {Deen}, \citenamefont {Florea},
  \citenamefont {Henry}, \citenamefont {Jacobsen}, \citenamefont {Knee},
  \citenamefont {Lhotel},\ and\ \citenamefont {Wildes}}]{D7_data2}%
  \BibitemOpen
  \bibfield  {author} {\bibinfo {author} {\bibfnamefont {P.~P.}\ \bibnamefont
  {Deen}}, \bibinfo {author} {\bibfnamefont {O.}~\bibnamefont {Florea}},
  \bibinfo {author} {\bibfnamefont {P.}~\bibnamefont {Henry}}, \bibinfo
  {author} {\bibfnamefont {H.}~\bibnamefont {Jacobsen}}, \bibinfo {author}
  {\bibfnamefont {C.}~\bibnamefont {Knee}}, \bibinfo {author} {\bibfnamefont
  {E.}~\bibnamefont {Lhotel}}, \ and\ \bibinfo {author} {\bibfnamefont
  {A.}~\bibnamefont {Wildes}},\ }\href@noop {} {\enquote {\bibinfo {title}
  {Emergent excitations and long range order in {Gd3Al5O12}},}\ }\bibinfo
  {howpublished} {Institut Laue-Langevin (ILL), doi:10.5291/ILL-DATA.5-32-795}
  (\bibinfo {year} {2014}{\natexlab{a}})\BibitemShut {NoStop}%
\bibitem [{\citenamefont {Ehlers}\ \emph {et~al.}(2013)\citenamefont {Ehlers},
  \citenamefont {Stewart}, \citenamefont {Wildes}, \citenamefont {Deen},\ and\
  \citenamefont {Andersen}}]{Ehlers2013}%
  \BibitemOpen
  \bibfield  {author} {\bibinfo {author} {\bibfnamefont {G.}~\bibnamefont
  {Ehlers}}, \bibinfo {author} {\bibfnamefont {J.~R.}\ \bibnamefont {Stewart}},
  \bibinfo {author} {\bibfnamefont {A.~R.}\ \bibnamefont {Wildes}}, \bibinfo
  {author} {\bibfnamefont {P.~P.}\ \bibnamefont {Deen}}, \ and\ \bibinfo
  {author} {\bibfnamefont {K.~H.}\ \bibnamefont {Andersen}},\ }\bibfield
  {title} {\enquote {\bibinfo {title} {{Generalization of the classical
  xyz-polarization analysis technique to out-of-plane and inelastic
  scattering}},}\ }\href {\doibase 10.1063/1.4819739} {\bibfield  {journal}
  {\bibinfo  {journal} {Review of Scientific Instruments}\ }\textbf {\bibinfo
  {volume} {84}},\ \bibinfo {pages} {093901} (\bibinfo {year}
  {2013})}\BibitemShut {NoStop}%
\bibitem [{\citenamefont {Stewart}\ \emph {et~al.}(2009)\citenamefont
  {Stewart}, \citenamefont {Deen}, \citenamefont {Andersen}, \citenamefont
  {Schober}, \citenamefont {Barth{\'{e}}l{\'{e}}my}, \citenamefont {Hillier},
  \citenamefont {Murani}, \citenamefont {Hayes},\ and\ \citenamefont
  {Lindenau}}]{Stewart2009}%
  \BibitemOpen
  \bibfield  {author} {\bibinfo {author} {\bibfnamefont {J.~R.}\ \bibnamefont
  {Stewart}}, \bibinfo {author} {\bibfnamefont {P.~P.}\ \bibnamefont {Deen}},
  \bibinfo {author} {\bibfnamefont {K.~H.}\ \bibnamefont {Andersen}}, \bibinfo
  {author} {\bibfnamefont {H.}~\bibnamefont {Schober}}, \bibinfo {author}
  {\bibfnamefont {J.}~\bibnamefont {Barth{\'{e}}l{\'{e}}my}}, \bibinfo {author}
  {\bibfnamefont {J.~M.}\ \bibnamefont {Hillier}}, \bibinfo {author}
  {\bibfnamefont {A.~P.}\ \bibnamefont {Murani}}, \bibinfo {author}
  {\bibfnamefont {T.}~\bibnamefont {Hayes}}, \ and\ \bibinfo {author}
  {\bibfnamefont {B.}~\bibnamefont {Lindenau}},\ }\bibfield  {title} {\enquote
  {\bibinfo {title} {{Disordered materials studied using neutron polarization
  analysis on the multi-detector spectrometer, D7}},}\ }\href {\doibase
  10.1107/S0021889808039162} {\bibfield  {journal} {\bibinfo  {journal} {J.
  Appl. Crystallogr.}\ }\textbf {\bibinfo {volume} {42}},\ \bibinfo {pages}
  {69} (\bibinfo {year} {2009})}\BibitemShut {NoStop}%
\bibitem [{\citenamefont {Richard}\ \emph {et~al.}(1996)\citenamefont
  {Richard}, \citenamefont {Ferrand},\ and\ \citenamefont
  {Kearley}}]{Richard:1996:J.NeutronRes.}%
  \BibitemOpen
  \bibfield  {author} {\bibinfo {author} {\bibfnamefont {D.}~\bibnamefont
  {Richard}}, \bibinfo {author} {\bibfnamefont {M.}~\bibnamefont {Ferrand}}, \
  and\ \bibinfo {author} {\bibfnamefont {G.~J.}\ \bibnamefont {Kearley}},\
  }\bibfield  {title} {\enquote {\bibinfo {title} {{Analysis and visualisation
  of neutron-scattering data}},}\ }\href {\doibase 10.1080/10238169608200065}
  {\bibfield  {journal} {\bibinfo  {journal} {Journal of Neutron Research}\
  }\textbf {\bibinfo {volume} {4}},\ \bibinfo {pages} {33} (\bibinfo {year}
  {1996})}\BibitemShut {NoStop}%
\bibitem [{\citenamefont
  {Rodr{\'{i}}guez-Carvajal}(1993)}]{Rodriguez-Carvajal1993}%
  \BibitemOpen
  \bibfield  {author} {\bibinfo {author} {\bibfnamefont {J.}~\bibnamefont
  {Rodr{\'{i}}guez-Carvajal}},\ }\bibfield  {title} {\enquote {\bibinfo {title}
  {{Recent advances in magnetic structure determination by neutron powder
  diffraction}},}\ }\href {\doibase 10.1016/0921-4526(93)90108-I} {\bibfield
  {journal} {\bibinfo  {journal} {Physica B: Condensed Matter}\ }\textbf
  {\bibinfo {volume} {192}},\ \bibinfo {pages} {55} (\bibinfo {year}
  {1993})}\BibitemShut {NoStop}%
\bibitem [{\citenamefont {Bewley}\ \emph {et~al.}(2011)\citenamefont {Bewley},
  \citenamefont {Taylor},\ and\ \citenamefont {Bennington.}}]{Bewley2011}%
  \BibitemOpen
  \bibfield  {author} {\bibinfo {author} {\bibfnamefont {R.~I.}\ \bibnamefont
  {Bewley}}, \bibinfo {author} {\bibfnamefont {J.~W.}\ \bibnamefont {Taylor}},
  \ and\ \bibinfo {author} {\bibfnamefont {S.~M.}\ \bibnamefont
  {Bennington.}},\ }\bibfield  {title} {\enquote {\bibinfo {title} {{LET, a
  cold neutron multi-disk chopper spectrometer at ISIS}},}\ }\href {\doibase
  10.1016/j.nima.2011.01.173} {\bibfield  {journal} {\bibinfo  {journal} {Nucl.
  Instrum. Methods Phys. Res. A}\ }\textbf {\bibinfo {volume} {637}},\ \bibinfo
  {pages} {128} (\bibinfo {year} {2011})}\BibitemShut {NoStop}%
\bibitem [{\citenamefont {Deen}\ \emph
  {et~al.}(2015{\natexlab{c}})\citenamefont {Deen}, \citenamefont {Lefmann},
  \citenamefont {Guyon-Le-Bouffy}, \citenamefont {Knee}, \citenamefont {Henry},
  \citenamefont {Petrenko}, \citenamefont {d'Ambrumenil}, \citenamefont
  {Jacobsen},\ and\ \citenamefont {Bewley}}]{LET_data}%
  \BibitemOpen
  \bibfield  {author} {\bibinfo {author} {\bibfnamefont {P.~P.}\ \bibnamefont
  {Deen}}, \bibinfo {author} {\bibfnamefont {K.}~\bibnamefont {Lefmann}},
  \bibinfo {author} {\bibfnamefont {J.}~\bibnamefont {Guyon-Le-Bouffy}},
  \bibinfo {author} {\bibfnamefont {C.}~\bibnamefont {Knee}}, \bibinfo {author}
  {\bibfnamefont {P.}~\bibnamefont {Henry}}, \bibinfo {author} {\bibfnamefont
  {O.}~\bibnamefont {Petrenko}}, \bibinfo {author} {\bibfnamefont
  {N.}~\bibnamefont {d'Ambrumenil}}, \bibinfo {author} {\bibfnamefont
  {H.}~\bibnamefont {Jacobsen}}, \ and\ \bibinfo {author} {\bibfnamefont
  {R.}~\bibnamefont {Bewley}},\ }\href@noop {} {\enquote {\bibinfo {title}
  {{Dispersionless spin waves and field-induced magnetic order in
  Gd3Al5O12}},}\ }\bibinfo {howpublished} {STFC ISIS Neutron and Muon Source,
  https://doi.org/10.5286/ISIS.E.RB1520308} (\bibinfo {year}
  {2015}{\natexlab{c}})\BibitemShut {NoStop}%
\bibitem [{\citenamefont {Arnold}\ \emph {et~al.}(2014)\citenamefont {Arnold},
  \citenamefont {Bilheux}, \citenamefont {Borreguero}, \citenamefont {Buts},
  \citenamefont {Campbell}, \citenamefont {Chapon}, \citenamefont {Doucet},
  \citenamefont {Draper}, \citenamefont {{Ferraz Leal}}, \citenamefont {Gigg},
  \citenamefont {Lynch}, \citenamefont {Markvardsen}, \citenamefont
  {Mikkelson}, \citenamefont {Mikkelson}, \citenamefont {Miller}, \citenamefont
  {Palmen}, \citenamefont {Parker}, \citenamefont {Passos}, \citenamefont
  {Perring}, \citenamefont {Peterson}, \citenamefont {Ren}, \citenamefont
  {Reuter}, \citenamefont {Savici}, \citenamefont {Taylor}, \citenamefont
  {Taylor}, \citenamefont {Tolchenov}, \citenamefont {Zhou},\ and\
  \citenamefont {Zikovsky}}]{Arnold2014}%
  \BibitemOpen
  \bibfield  {author} {\bibinfo {author} {\bibfnamefont {O.}~\bibnamefont
  {Arnold}}, \bibinfo {author} {\bibfnamefont {J.~C.}\ \bibnamefont {Bilheux}},
  \bibinfo {author} {\bibfnamefont {J.~M.}\ \bibnamefont {Borreguero}},
  \bibinfo {author} {\bibfnamefont {A.}~\bibnamefont {Buts}}, \bibinfo {author}
  {\bibfnamefont {S.~I.}\ \bibnamefont {Campbell}}, \bibinfo {author}
  {\bibfnamefont {L.}~\bibnamefont {Chapon}}, \bibinfo {author} {\bibfnamefont
  {M.}~\bibnamefont {Doucet}}, \bibinfo {author} {\bibfnamefont
  {N.}~\bibnamefont {Draper}}, \bibinfo {author} {\bibfnamefont
  {R.}~\bibnamefont {{Ferraz Leal}}}, \bibinfo {author} {\bibfnamefont {M.~A.}\
  \bibnamefont {Gigg}}, \bibinfo {author} {\bibfnamefont {V.~E.}\ \bibnamefont
  {Lynch}}, \bibinfo {author} {\bibfnamefont {A.}~\bibnamefont {Markvardsen}},
  \bibinfo {author} {\bibfnamefont {D.~J.}\ \bibnamefont {Mikkelson}}, \bibinfo
  {author} {\bibfnamefont {R.~L.}\ \bibnamefont {Mikkelson}}, \bibinfo {author}
  {\bibfnamefont {R.}~\bibnamefont {Miller}}, \bibinfo {author} {\bibfnamefont
  {K.}~\bibnamefont {Palmen}}, \bibinfo {author} {\bibfnamefont
  {P.}~\bibnamefont {Parker}}, \bibinfo {author} {\bibfnamefont
  {G.}~\bibnamefont {Passos}}, \bibinfo {author} {\bibfnamefont {T.~G.}\
  \bibnamefont {Perring}}, \bibinfo {author} {\bibfnamefont {P.~F.}\
  \bibnamefont {Peterson}}, \bibinfo {author} {\bibfnamefont {S.}~\bibnamefont
  {Ren}}, \bibinfo {author} {\bibfnamefont {M.~A.}\ \bibnamefont {Reuter}},
  \bibinfo {author} {\bibfnamefont {A.~T.}\ \bibnamefont {Savici}}, \bibinfo
  {author} {\bibfnamefont {J.~W.}\ \bibnamefont {Taylor}}, \bibinfo {author}
  {\bibfnamefont {R.~J.}\ \bibnamefont {Taylor}}, \bibinfo {author}
  {\bibfnamefont {R.}~\bibnamefont {Tolchenov}}, \bibinfo {author}
  {\bibfnamefont {W.}~\bibnamefont {Zhou}}, \ and\ \bibinfo {author}
  {\bibfnamefont {J.}~\bibnamefont {Zikovsky}},\ }\bibfield  {title} {\enquote
  {\bibinfo {title} {{Mantid - Data analysis and visualization package for
  neutron scattering and $\mu$SR experiments}},}\ }\href {\doibase
  10.1016/j.nima.2014.07.029} {\bibfield  {journal} {\bibinfo  {journal} {Nucl.
  Instrum. Methods Phys. Res. A}\ }\textbf {\bibinfo {volume} {764}},\ \bibinfo
  {pages} {156} (\bibinfo {year} {2014})}\BibitemShut {NoStop}%
\bibitem [{\citenamefont {Deen}\ \emph
  {et~al.}(2014{\natexlab{b}})\citenamefont {Deen}, \citenamefont {Frick},
  \citenamefont {Jacobsen},\ and\ \citenamefont {Seydel}}]{IN16b_data}%
  \BibitemOpen
  \bibfield  {author} {\bibinfo {author} {\bibfnamefont {P.~P.}\ \bibnamefont
  {Deen}}, \bibinfo {author} {\bibfnamefont {B.}~\bibnamefont {Frick}},
  \bibinfo {author} {\bibfnamefont {H.}~\bibnamefont {Jacobsen}}, \ and\
  \bibinfo {author} {\bibfnamefont {T.}~\bibnamefont {Seydel}},\ }\href@noop {}
  {\enquote {\bibinfo {title} {{ Exploring microeV dynamics in {Gadolinium
  Gallium Garnet}}},}\ }\bibinfo {howpublished} {Institut Laue-Langevin (ILL),
  doi:10.5291/ILL-DATA.4-03-1704} (\bibinfo {year}
  {2014}{\natexlab{b}})\BibitemShut {NoStop}%
\bibitem [{\citenamefont {Frick}\ \emph {et~al.}(2010)\citenamefont {Frick},
  \citenamefont {Mamontov}, \citenamefont {Eijck},\ and\ \citenamefont
  {Seydel}}]{Frick2010}%
  \BibitemOpen
  \bibfield  {author} {\bibinfo {author} {\bibfnamefont {B.}~\bibnamefont
  {Frick}}, \bibinfo {author} {\bibfnamefont {E.}~\bibnamefont {Mamontov}},
  \bibinfo {author} {\bibfnamefont {L.~V.}\ \bibnamefont {Eijck}}, \ and\
  \bibinfo {author} {\bibfnamefont {T.}~\bibnamefont {Seydel}},\ }\bibfield
  {title} {\enquote {\bibinfo {title} {{Recent Backscattering Instrument
  Developments at the ILL and SNS}},}\ }\href {\doibase 10.1524/zpch.2010.6091}
  {\bibfield  {journal} {\bibinfo  {journal} {Zeitschrift f{\"{u}}r
  Physikalische Chemie}\ }\textbf {\bibinfo {volume} {224}},\ \bibinfo {pages}
  {33} (\bibinfo {year} {2010})}\BibitemShut {NoStop}%
\bibitem [{\citenamefont {Paulsen}()}]{ExtractionMethod}%
  \BibitemOpen
  \bibfield  {author} {\bibinfo {author} {\bibfnamefont {C.}~\bibnamefont
  {Paulsen}},\ }\href@noop {} {\enquote {\bibinfo {title} {Introduction to
  physical techniques in molecular magnetism: Structural and macroscopic
  techniques},}\ }\bibinfo {howpublished} {(University of Zaragoza, Zaragoza,
  2001)}\BibitemShut {NoStop}%
\bibitem [{\citenamefont {Paddison}\ \emph {et~al.}(2013)\citenamefont
  {Paddison}, \citenamefont {{Ross Stewart}},\ and\ \citenamefont
  {Goodwin}}]{Paddison2013a}%
  \BibitemOpen
  \bibfield  {author} {\bibinfo {author} {\bibfnamefont {J.~A.~M.}\
  \bibnamefont {Paddison}}, \bibinfo {author} {\bibfnamefont {J.}~\bibnamefont
  {{Ross Stewart}}}, \ and\ \bibinfo {author} {\bibfnamefont {A.~L.}\
  \bibnamefont {Goodwin}},\ }\bibfield  {title} {\enquote {\bibinfo {title}
  {{Spinvert: a Program for Refinement of Paramagnetic Diffuse Scattering
  Data}},}\ }\href {\doibase 10.1088/0953-8984/25/45/454220} {\bibfield
  {journal} {\bibinfo  {journal} {J. Phys. Condens. Matter}\ }\textbf {\bibinfo
  {volume} {25}},\ \bibinfo {pages} {454220} (\bibinfo {year}
  {2013})}\BibitemShut {NoStop}%
\bibitem [{\citenamefont {Metropolis}\ and\ \citenamefont
  {Ulam}(1949)}]{Metropolis1949}%
  \BibitemOpen
  \bibfield  {author} {\bibinfo {author} {\bibfnamefont {N.}~\bibnamefont
  {Metropolis}}\ and\ \bibinfo {author} {\bibfnamefont {S.}~\bibnamefont
  {Ulam}},\ }\bibfield  {title} {\enquote {\bibinfo {title} {{The Monte Carlo
  Method}},}\ }\href {\doibase 10.1080/01621459.1949.10483310} {\bibfield
  {journal} {\bibinfo  {journal} {Journal of the American Statistical
  Association}\ }\textbf {\bibinfo {volume} {44}},\ \bibinfo {pages} {335}
  (\bibinfo {year} {1949})}\BibitemShut {NoStop}%
\bibitem [{\citenamefont {Mukherjee}\ \emph {et~al.}(2017)\citenamefont
  {Mukherjee}, \citenamefont {Hamilton}, \citenamefont {Glass},\ and\
  \citenamefont {Dutton}}]{Mukherjee2017}%
  \BibitemOpen
  \bibfield  {author} {\bibinfo {author} {\bibfnamefont {P.}~\bibnamefont
  {Mukherjee}}, \bibinfo {author} {\bibfnamefont {A.~C.~S.}\ \bibnamefont
  {Hamilton}}, \bibinfo {author} {\bibfnamefont {H.~F.~J.}\ \bibnamefont
  {Glass}}, \ and\ \bibinfo {author} {\bibfnamefont {S.~E.}\ \bibnamefont
  {Dutton}},\ }\bibfield  {title} {\enquote {\bibinfo {title} {{Sensitivity of
  magnetic properties to chemical pressure in lanthanide garnets
  Ln$_3$A$_2$X$_3$O$_{12}$, Ln = Gd, Tb, Dy, Ho, A = Ga, Sc, In, Te, X = Ga,
  Al, Li}},}\ }\href@noop {} {\bibfield  {journal} {\bibinfo  {journal} {J.
  Phys. Condens. Matter}\ }\textbf {\bibinfo {volume} {29}},\ \bibinfo {pages}
  {405908} (\bibinfo {year} {2017})}\BibitemShut {NoStop}%
\bibitem [{\citenamefont {Sandberg}\ \emph {et~al.}(2020)\citenamefont
  {Sandberg}, \citenamefont {Edberg}, \citenamefont {Bakke}, \citenamefont
  {Pedersen}, \citenamefont {Hatnean}, \citenamefont {Balaskrishnan},
  \citenamefont {Mangin-Thro}, \citenamefont {Wildes}, \citenamefont {F{\aa}k},
  \citenamefont {Ehlers}, \citenamefont {Sala}, \citenamefont {Henelius},
  \citenamefont {Lefmann},\ and\ \citenamefont {Deen}}]{Sandberg2020}%
  \BibitemOpen
  \bibfield  {author} {\bibinfo {author} {\bibfnamefont {L.~\O{}.}\
  \bibnamefont {Sandberg}}, \bibinfo {author} {\bibfnamefont {R.}~\bibnamefont
  {Edberg}}, \bibinfo {author} {\bibfnamefont {I.~M.~B.}\ \bibnamefont
  {Bakke}}, \bibinfo {author} {\bibfnamefont {K.~S.}\ \bibnamefont {Pedersen}},
  \bibinfo {author} {\bibfnamefont {M.~C.}\ \bibnamefont {Hatnean}}, \bibinfo
  {author} {\bibfnamefont {G.}~\bibnamefont {Balaskrishnan}}, \bibinfo {author}
  {\bibfnamefont {L.}~\bibnamefont {Mangin-Thro}}, \bibinfo {author}
  {\bibfnamefont {A.}~\bibnamefont {Wildes}}, \bibinfo {author} {\bibfnamefont
  {B.}~\bibnamefont {F{\aa}k}}, \bibinfo {author} {\bibfnamefont
  {G.}~\bibnamefont {Ehlers}}, \bibinfo {author} {\bibfnamefont
  {G.}~\bibnamefont {Sala}}, \bibinfo {author} {\bibfnamefont {P.}~\bibnamefont
  {Henelius}}, \bibinfo {author} {\bibfnamefont {K.}~\bibnamefont {Lefmann}}, \
  and\ \bibinfo {author} {\bibfnamefont {P.~P.}\ \bibnamefont {Deen}},\
  }\bibfield  {title} {\enquote {\bibinfo {title} {{Emergent magnetic behavior
  in the frustrated Yb3Ga5O12 garnet}},}\ }\href@noop {} {\  (\bibinfo {year}
  {2020})},\ \Eprint {http://arxiv.org/abs/2005.10605} {arXiv:2005.10605}
  \BibitemShut {NoStop}%
\bibitem [{\citenamefont {Boothroyd}(2020)}]{Boothroyd2020}%
  \BibitemOpen
  \bibfield  {author} {\bibinfo {author} {\bibfnamefont {A.}~\bibnamefont
  {Boothroyd}},\ }\href@noop {} {\emph {\bibinfo {title} {Principles of Neutron
  Scattering from Condensed Matter}}}\ (\bibinfo  {publisher} {Oxford
  University Press},\ \bibinfo {address} {Oxford},\ \bibinfo {year}
  {2020})\BibitemShut {NoStop}%
\bibitem [{\citenamefont {Mirebeau}\ \emph {et~al.}(2008)\citenamefont
  {Mirebeau}, \citenamefont {Mutka}, \citenamefont {Bonville}, \citenamefont
  {Apetrei},\ and\ \citenamefont {Forget}}]{Mirebeau2008}%
  \BibitemOpen
  \bibfield  {author} {\bibinfo {author} {\bibfnamefont {I.}~\bibnamefont
  {Mirebeau}}, \bibinfo {author} {\bibfnamefont {H.}~\bibnamefont {Mutka}},
  \bibinfo {author} {\bibfnamefont {P.}~\bibnamefont {Bonville}}, \bibinfo
  {author} {\bibfnamefont {A.}~\bibnamefont {Apetrei}}, \ and\ \bibinfo
  {author} {\bibfnamefont {A.}~\bibnamefont {Forget}},\ }\bibfield  {title}
  {\enquote {\bibinfo {title} {{Investigation of magnetic fluctuations in
  Tb$_2$Sn$_2$O$_7$}},}\ }\href {\doibase 10.1103/PhysRevB.78.174416}
  {\bibfield  {journal} {\bibinfo  {journal} {Phys. Rev. B}\ }\textbf {\bibinfo
  {volume} {78}},\ \bibinfo {pages} {174416} (\bibinfo {year}
  {2008})}\BibitemShut {NoStop}%
\bibitem [{\citenamefont {Conlon}\ and\ \citenamefont
  {Chalker}(2009)}]{Conlon2009}%
  \BibitemOpen
  \bibfield  {author} {\bibinfo {author} {\bibfnamefont {P.~H.}\ \bibnamefont
  {Conlon}}\ and\ \bibinfo {author} {\bibfnamefont {J.~T.}\ \bibnamefont
  {Chalker}},\ }\bibfield  {title} {\enquote {\bibinfo {title} {{Spin Dynamics
  in Pyrochlore Heisenberg Antiferromagnets}},}\ }\href {\doibase
  10.1103/PhysRevLett.102.237206} {\bibfield  {journal} {\bibinfo  {journal}
  {Phys. Rev. Lett.}\ }\textbf {\bibinfo {volume} {102}},\ \bibinfo {pages}
  {237206} (\bibinfo {year} {2009})}\BibitemShut {NoStop}%
\bibitem [{\citenamefont {Taillefumier}\ \emph {et~al.}(2014)\citenamefont
  {Taillefumier}, \citenamefont {Robert}, \citenamefont {Henley}, \citenamefont
  {Moessner},\ and\ \citenamefont {Canals}}]{Taillefumier2014}%
  \BibitemOpen
  \bibfield  {author} {\bibinfo {author} {\bibfnamefont {M.}~\bibnamefont
  {Taillefumier}}, \bibinfo {author} {\bibfnamefont {J.}~\bibnamefont
  {Robert}}, \bibinfo {author} {\bibfnamefont {C.~L.}\ \bibnamefont {Henley}},
  \bibinfo {author} {\bibfnamefont {R.}~\bibnamefont {Moessner}}, \ and\
  \bibinfo {author} {\bibfnamefont {B.}~\bibnamefont {Canals}},\ }\bibfield
  {title} {\enquote {\bibinfo {title} {{Semiclassical spin dynamics of the
  antiferromagnetic Heisenberg model on the kagome lattice}},}\ }\href
  {\doibase 10.1103/PhysRevB.90.064419} {\bibfield  {journal} {\bibinfo
  {journal} {Phys. Rev. B}\ }\textbf {\bibinfo {volume} {90}},\ \bibinfo
  {pages} {064419} (\bibinfo {year} {2014})}\BibitemShut {NoStop}%
\bibitem [{\citenamefont {Topping}\ and\ \citenamefont
  {Blundell}(2019)}]{Topping2019}%
  \BibitemOpen
  \bibfield  {author} {\bibinfo {author} {\bibfnamefont {C.~V.}\ \bibnamefont
  {Topping}}\ and\ \bibinfo {author} {\bibfnamefont {S.~J.}\ \bibnamefont
  {Blundell}},\ }\bibfield  {title} {\enquote {\bibinfo {title} {{A.C.
  susceptibility as a probe of low-frequency magnetic dynamics}},}\ }\href
  {\doibase 10.1088/1361-648X/aaed96} {\bibfield  {journal} {\bibinfo
  {journal} {J. Phys. Condens. Matter}\ }\textbf {\bibinfo {volume} {31}},\
  \bibinfo {pages} {013001} (\bibinfo {year} {2019})}\BibitemShut {NoStop}%
\bibitem [{\citenamefont {Schiffer}\ \emph {et~al.}(1995)\citenamefont
  {Schiffer}, \citenamefont {Ramirez}, \citenamefont {Huse}, \citenamefont
  {Gammel}, \citenamefont {Yaron}, \citenamefont {Bishop},\ and\ \citenamefont
  {Valentino}}]{Schiffer1995}%
  \BibitemOpen
  \bibfield  {author} {\bibinfo {author} {\bibfnamefont {P.}~\bibnamefont
  {Schiffer}}, \bibinfo {author} {\bibfnamefont {A.~P.}\ \bibnamefont
  {Ramirez}}, \bibinfo {author} {\bibfnamefont {D.~A.}\ \bibnamefont {Huse}},
  \bibinfo {author} {\bibfnamefont {P.~L.}\ \bibnamefont {Gammel}}, \bibinfo
  {author} {\bibfnamefont {U.}~\bibnamefont {Yaron}}, \bibinfo {author}
  {\bibfnamefont {D.~J.}\ \bibnamefont {Bishop}}, \ and\ \bibinfo {author}
  {\bibfnamefont {A.~J.}\ \bibnamefont {Valentino}},\ }\bibfield  {title}
  {\enquote {\bibinfo {title} {{Frustration Induced Spin Freezing in a
  Site-Ordered magnet: Gadolinium Gallium Garnet}},}\ }\href {\doibase
  10.1103/PhysRevLett.74.2379} {\bibfield  {journal} {\bibinfo  {journal}
  {Phys. Rev. Lett.}\ }\textbf {\bibinfo {volume} {74}},\ \bibinfo {pages}
  {2379} (\bibinfo {year} {1995})}\BibitemShut {NoStop}%
\bibitem [{\citenamefont {Shtrikman}\ and\ \citenamefont
  {Wohlfarth}(1981)}]{Shtrikman1981}%
  \BibitemOpen
  \bibfield  {author} {\bibinfo {author} {\bibfnamefont {S.}~\bibnamefont
  {Shtrikman}}\ and\ \bibinfo {author} {\bibfnamefont {E.P.}\ \bibnamefont
  {Wohlfarth}},\ }\bibfield  {title} {\enquote {\bibinfo {title} {The theory of
  the vogel-fulcher law of spin glasses},}\ }\href {\doibase
  https://doi.org/10.1016/0375-9601(81)90441-2} {\bibfield  {journal} {\bibinfo
   {journal} {Physics Letters A}\ }\textbf {\bibinfo {volume} {85}},\ \bibinfo
  {pages} {467--470} (\bibinfo {year} {1981})}\BibitemShut {NoStop}%
\bibitem [{\citenamefont {Mydosh}(1993)}]{Mydosh1993}%
  \BibitemOpen
  \bibfield  {author} {\bibinfo {author} {\bibfnamefont {J.}~\bibnamefont
  {Mydosh}},\ }\href@noop {} {\emph {\bibinfo {title} {Spin Glasses: An
  Experimental Introduction (1st ed.)}}}\ (\bibinfo  {publisher} {CRC Press},\
  \bibinfo {year} {1993})\ \bibinfo {note}
  {https://doi.org/10.1201/9781482295191}\BibitemShut {NoStop}%
\bibitem [{\citenamefont {C{\'{e}}pas}\ and\ \citenamefont
  {Canals}(2012)}]{Cepas2012}%
  \BibitemOpen
  \bibfield  {author} {\bibinfo {author} {\bibfnamefont {O.}~\bibnamefont
  {C{\'{e}}pas}}\ and\ \bibinfo {author} {\bibfnamefont {B.}~\bibnamefont
  {Canals}},\ }\bibfield  {title} {\enquote {\bibinfo {title} {{Heterogeneous
  freezing in a geometrically frustrated spin model without disorder:
  Spontaneous generation of two time scales}},}\ }\href {\doibase
  10.1103/PhysRevB.86.024434} {\bibfield  {journal} {\bibinfo  {journal} {Phys.
  Rev. B}\ }\textbf {\bibinfo {volume} {86}},\ \bibinfo {pages} {024434}
  (\bibinfo {year} {2012})}\BibitemShut {NoStop}%
\bibitem [{\citenamefont {Beauvillain}\ \emph {et~al.}(1984)\citenamefont
  {Beauvillain}, \citenamefont {Dupas}, \citenamefont {Renard},\ and\
  \citenamefont {Veillet}}]{Beauvillain1984}%
  \BibitemOpen
  \bibfield  {author} {\bibinfo {author} {\bibfnamefont {P.}~\bibnamefont
  {Beauvillain}}, \bibinfo {author} {\bibfnamefont {C.}~\bibnamefont {Dupas}},
  \bibinfo {author} {\bibfnamefont {J.~P.}\ \bibnamefont {Renard}}, \ and\
  \bibinfo {author} {\bibfnamefont {P.}~\bibnamefont {Veillet}},\ }\bibfield
  {title} {\enquote {\bibinfo {title} {{Experimental study of the spin freezing
  in an insulating spin-glass: Static and dynamical aspects}},}\ }\href
  {\doibase 10.1103/PhysRevB.29.4086} {\bibfield  {journal} {\bibinfo
  {journal} {Phys. Rev. B}\ }\textbf {\bibinfo {volume} {29}},\ \bibinfo
  {pages} {4086} (\bibinfo {year} {1984})}\BibitemShut {NoStop}%
\end{thebibliography}
%

\clearpage
\appendix

\section{Analysis of backscattering data}

\begin{figure}[ht]
 \centering
\includegraphics[width=0.23\textwidth]{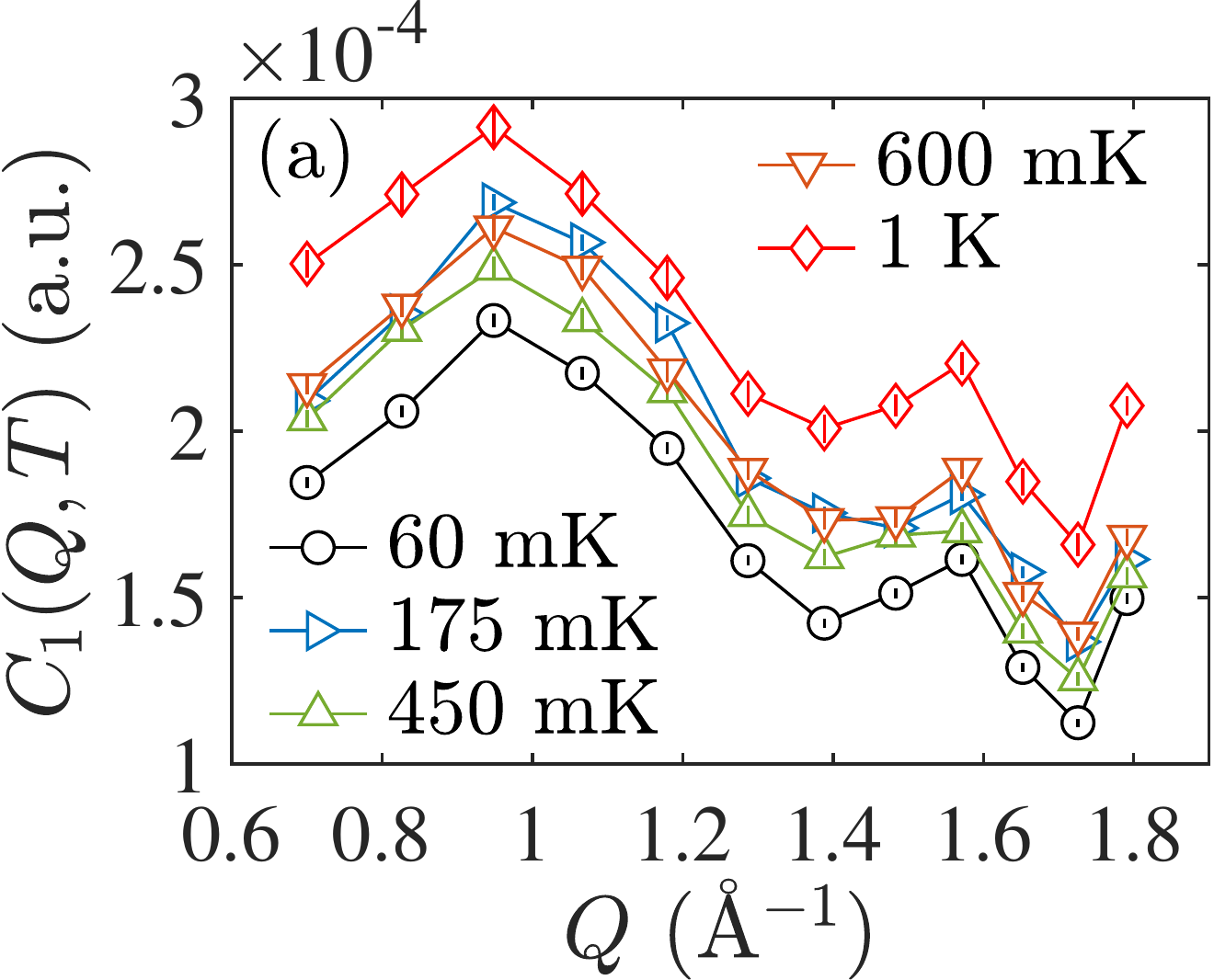}
 \includegraphics[width=0.23\textwidth]{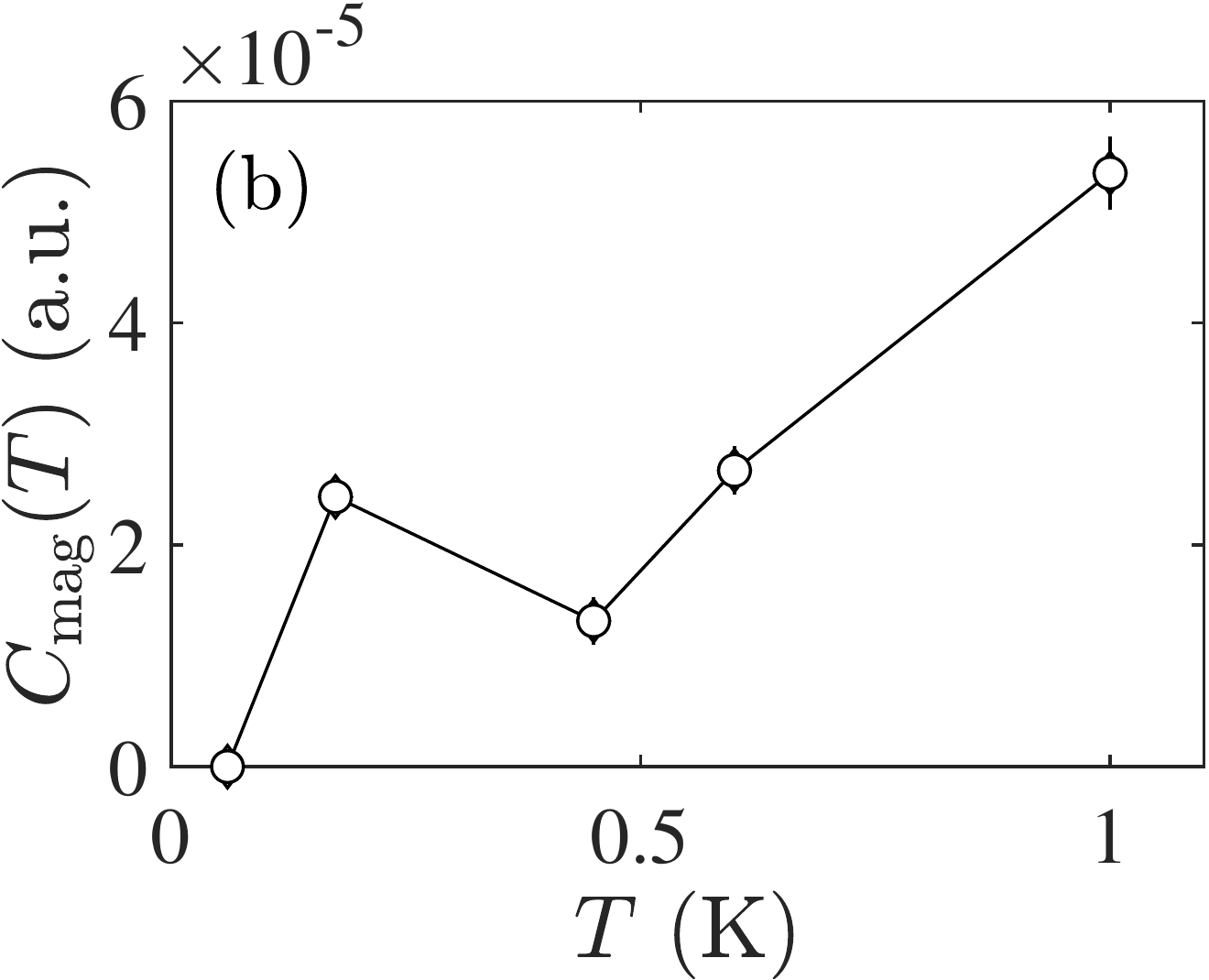}
\caption{The flat background signal of GGG measured at IN16b. (a) shows the full signal,   (b) shows the magnetic signal, $C_\text{mag}(T)$.}
  \label{fig:Fig12}
 \end{figure}

The Lorentzian signal sits atop a linear background, with the flat part $C_1(Q,T)$, increasing with temperature as shown in Fig.~\ref{fig:Fig12}(a). The slope of the background depends slightly on $Q$ and shows no systematic temperature dependence. Importantly, the slope is small compared with the flat part, and we can ignore it in the following analysis.
The flat part of the background signal depends on temperature and $Q$, but can be subdivided into two parts to represent a background contribution from sample environment ($T$-independent part $(C_\text{env}(Q))$) and a magnetic contribution, (Q-independent part, $C_\text{mag}(T)$):
\begin{align}
 C_1(Q,T)=C_\text{env}(Q)+C_\text{mag}(T).
\end{align}
Fig.~\ref{fig:Fig12}(b) shows $C_\text{mag}(T)$, the magnetic component. 

$C_\text{mag}(T)$ increases with increasing temperature and thus indicates that it originates from motions on a larger energy scale than the accessible energy window of IN16b ($\sim \pm 30$~$\mu$eV), similar to previous backscattering measurements on Tb$_2$Sn$_2$O$_7$ \cite{Mirebeau2008}. It is likely that this contribution originates from the tails of the lowest excitation, INS1, observed at 0.05(1) meV. 

 \end{document}